\documentclass[reprint,superscriptaddress,secnumarabic,floatfix,amsmath,amssymb, nobibnotes,aps,prb]{revtex4-2}
\usepackage{graphicx}
\usepackage{dcolumn}
\usepackage{amsmath}

\usepackage{bm}
\usepackage{dcolumn}
\usepackage{varioref}
\usepackage[dvipsnames]{xcolor}
\usepackage{hyperref}
\usepackage[capitalize]{cleveref}

\usepackage{bm}
\usepackage{dcolumn}
\usepackage{amsmath}
\usepackage{graphicx}

\usepackage{dcolumn}
\usepackage{bm}
\usepackage[margin=0.4in]{geometry}
\usepackage{multirow}
\usepackage{xcolor}
\usepackage[range-units=single,per-mode=reciprocal]{siunitx}

\usepackage{todonotes}

\usepackage[version=4]{mhchem}
\DeclareSIUnit\rydberg{Ry}
\begin{document}
	\title{
    Spectroscopy and transport of nonpolarons in silicon and germanium: the influence of doping and temperature}
	\author{Raveena Gupta}
    \affiliation{Nanomat group, Q-MAT, University of Liège,  and European Theoretical Spectroscopy Facility, B-4000 Liège, Belgium}
    \author{Joao Abreu}
    \affiliation{Nanomat group, Q-MAT, University of Liège,  and European Theoretical Spectroscopy Facility, B-4000 Liège, Belgium}	
    \author{Matthieu J. Verstraete}
	\email{matthieu.verstraete@uliege.be}
    \affiliation{Nanomat group, Q-MAT, University of Liège,  and European Theoretical Spectroscopy Facility, B-4000 Liège, Belgium}%
    \affiliation{ITP, Physics Department, Utrecht University 3508 TA Utrecht, The Netherlands}
	\begin{abstract}
		We perform a first-principles investigation of electron–phonon interactions in silicon and germanium, uncovering distinct non-polaronic spectral and transport fingerprints in these archetypal covalent semiconductors. Using many-body perturbation theory with the retarded cumulant expansion, we compute quasiparticle energies, lifetimes, and phonon satellites beyond the Dyson–Migdal approximation. Short-range crystal fields dominate coupling in both materials, yet their low-temperature spectral fingerprints differ: Si exhibits well-resolved satellites at both band edges, whereas Ge displays strong sidebands mainly at the valence band maximum (VBM) and much weaker features at the conduction band minimum (CBM). Phonon-induced satellites in both materials broaden and merge with the quasiparticle peak at elevated temperatures. Doping broadens peaks and compresses satellite–quasiparticle separation, with n-type carriers affecting the CBM and p-type the VBM. Mobility calculations, combining cumulant-derived phonon scattering with experimentally motivated ionized-impurity scattering models, reproduce measured trends and reveal Ge’s consistently higher mobilities than Si, stemming from lighter effective masses and weaker coupling. These results link band-edge asymmetries and phonon energetics to measurable transport differences, providing a unified framework for predicting mobility in nonpolar semiconductors. 
	\end{abstract}
	
	\maketitle
	\section{INTRODUCTION}

The study of electron-phonon (e-ph) interactions lies at the heart of modern condensed matter physics due to its profound impact on both fundamental mechanisms (e.g. superconductivity, thermo and magneto transport) and on practical device performance (e.g. the temperature and doping dependency of semiconductor mobility). Silicon and germanium are the archetypal semiconductors, both covalent materials with tetrahedral bonding structures.  These materials exhibit relatively high mobilities for both electrons and holes, which underpins their widespread use in microelectronics. 
E-ph interactions play a decisive role in determining their electronic transport\cite{jacoboni1989monte, rode1975low, canali1975electron,gerDebye1952,gerJohnson1950,gerKoenig1957,gerKoenig1962,gerMorin1954}, particularly under conditions of low carrier concentration, elevated temperatures, or strong electric fields.\cite{pop2010heat, nayfeh2005effects} As device dimensions continue to scale down and performance demands increase, an accurate description of e-ph interactions becomes increasingly critical for predictive modeling of transport and optical responses.

In polar materials, e-ph coupling can lead to the formation of polarons-quasiparticles where electrons are dressed by lattice distortions.\cite{frohlich1954electrons} In contrast, silicon and germanium lack long-range dipole fields due to their purely covalent nature.  
Polaronic sidebands have been observed in diamond, another covalent material, suggesting that polaron-like signatures could also be present in silicon and germanium, arising from strong but non-dipolar e-ph coupling, for which Emin coined the term nonpolaron in Ref.~\cite{emin1989polaron}. 
A natural extension would lead to polaron-like behavior via interactions with long-range multipole fields, such as quadrupolar interactions, but we have shown previously that, at least in the case of diamond\cite{abreu2022}, the quadrupole contribution is minor and more localized fields are the dominant source of e-ph scattering.
Understanding these effects is essential for unraveling the microscopic origins of transport and spectral broadening in such systems.

The one-electron spectral function \(A(\omega)\) serves as a powerful tool for probing quasiparticle (QP) dynamics and many-body interactions. Angle-resolved photoemission spectroscopy (ARPES), in particular, provides direct experimental access to the spectral function, capturing key signatures such as QP energies, lifetimes (via linewidths), and satellite features associated with phonon coupling. Modeling the spectral function with high fidelity requires going beyond standard approximations. While the Migdal approximation\cite{Migdal,Dyson} provides a first-order perturbative treatment of e-ph coupling, it often fails to capture the correct energy positions and intensities of phonon satellites.\cite{guzzo2012, caruso2015, kas2014cumulant} 

The cumulant expansion (CE) formalism\cite{aryasetiawan1996, gumhalter2016cumulant} offers a more accurate theoretical framework by systematically including higher-order e-ph processes and multi-phonon excitations. It has been shown to significantly improve agreement with experiment in materials where many-body interactions are strong or particle-hole asymmetry is present.\cite{nery2018quasiparticles} For non-polar semiconductors like silicon and germanium, the cumulant expansion is particularly appealing, as it allows one to assess the subtle consequences of nonpolaronic interactions on the spectral and transport properties.

In this work, we employ the retarded CE formalism to study the spectral functions of both silicon and germanium, focusing on temperature-dependent renormalization of electronic energies, QP broadening, and the manifestation of nonpolaronic sidebands. To calculate transport and spectra on the same footing, we employ the Kubo Greenwood formula for the electrical conductivity and mobility. We further explore the effects of carrier doping---both \(n\)- and \(p\)-type---on the e-ph coupling strength and spectral features, and evaluate the implications for carrier mobility.  By comparing our theoretical predictions with experimental benchmarks, we aim to provide a comprehensive understanding of how e-ph interactions shape the electronic and transport behavior in these two cornerstone semiconductors. This study not only captures subtle nonpolaronic signatures beyond the reach of standard Migdal theory but also helps bridge the gap between fundamental many-body theory and technologically relevant materials.
	
\section{METHODOLOGY}

In this study, we investigate e–ph interactions using Many-Body Perturbation Theory (MBPT) techniques built upon ingredients from Density Functional Theory (DFT) and Density Functional Perturbation Theory (DFPT). While MBPT provides a robust framework for describing weak to moderate coupling regimes, it may become inadequate in systems where strong coupling leads to the formation of small, localized polarons—conditions under which perturbative treatments break down. The underlying computational details are briefly summarized in \cref{appendix:computational_detail}.

\subsection{Self energy}
We employ the MBPT formalism to detail the e-ph interaction. Specifically, we start by focussing on the lowest order e-ph self-energy,\cite{nery2018quasiparticles} also known as the Fan-Migdal (FM) self-energy, which is second order in atomic displacements. The FM self-energy comprises two components: the static Debye-Waller (DW) term and the dynamic Fan term and is given by,\cite{verstraete2024polaronic} 
\begin{equation}
\Sigma_{n\mathbf{k}}(\omega) = \Sigma^{DW}_{n\mathbf{k}} + \Sigma^{Fan}_{n\mathbf{k}}(\omega) 
\label{eq:1_self_energy_DW+FAN}
\end{equation}
where
\begin{equation}
\Sigma^{DW}_{n\mathbf{k}} = \frac{1}{N_\mathbf{q}} \sum_{j\mathbf{q}} g^{DW,j\mathbf{q},j-\mathbf{q}}_{nn\mathbf{k}} [2n_{j\mathbf{q}}(T) + 1] 
\label{eq:2_DW_self_energy}
\end{equation}
and

\begin{equation}
	\begin{split}
		\Sigma^{Fan}_{n\mathbf{k}}(\omega) = \frac{1}{N_\mathbf{q}} \sum_{m} \sum_{j\mathbf{q}} |g^{j\mathbf{q}}_{nm\mathbf{k}}|^2 
		&\left[ 
		\frac{n_{j\mathbf{q}}(T) + f_{m\mathbf{k}+\mathbf{q}}(T)}{\omega - \epsilon_{m\mathbf{k}+\mathbf{q}} + \omega_{j\mathbf{q}} + i\zeta} + 
		\right.\\
		&\left.
		\frac{n_{j\mathbf{q}}(T) + 1 - f_{m\mathbf{k}+\mathbf{q}}(T)}{\omega - \epsilon_{m\mathbf{k}+\mathbf{q}} - \omega_{j\mathbf{q}} + i\zeta} 
		\right]
	\end{split}
	\label{eq:3_Fan_self_energy}
\end{equation}

In these equations, \( \eta \) is a positive infinitesimal, \( \epsilon_{m\mathbf{k}} \) denotes the energy of the electronic state with band index \( m \) and wavevector \( \mathbf{k} \), and \( f_{m\mathbf{k}} \) is the Fermi-Dirac occupation number. The phonon frequencies are denoted as \( \omega_{j\mathbf{q}} \), with \( j \) being the mode index and \( \mathbf{q} \) the phonon wave vector. The Bose-Einstein occupation number is represented by \( n_{j\mathbf{q}} \), and \( g^{j\mathbf{q}}_{nm\mathbf{k}} \) is the first-order e-ph matrix element.  We use the acoustic sum rule and the rigid-ion approximation to express \( g^{DW} \) in terms of the first-order e-ph matrix elements. The terms inside the parenthesis of Eq. (3) represent absorption and emission of phonons, pairing with the appropriate electron
or hole state, respectively.

\subsection{Spectral function}
In general, the electron spectral function near the Fermi level is characterized by a sharp QP peak, accompanied by satellites due to phonon interactions. However, in polar materials,\cite{nery2018quasiparticles} the Dyson-Migdal (DM)\cite{Dyson,Migdal} approach often produces inaccurate QP energies and spectral weights compared to high-quality Monte Carlo results for the Fröhlich model.\cite{Mishchenko2000} The DM method also tends to misplace the satellite relative to the QP peak, which ideally should align with the phonon frequency \( \omega_{LO} \) of the LO mode. To improve upon the limitations of the DM approximation, we explore the CE method.\cite{Kubo1962,Gunnarsson1994} Herein, we compare the standard Dyson equation in the Migdal approximation with the CE method to compute spectral functions and QP energies. The details of both approaches are outlined in the following subsections. 

\subsubsection{Dyson-Migdal framework}

According to Migdal’s theorem,\cite{Migdal} the leading-order contributions of the e-ph interaction to the self-energy are the most significant due to the substantial mass difference between electrons and nuclei. Consequently, the electron self-energy can be approximated using the simplest diagram, allowing vertex corrections to be neglected. The electron Green's function \( G_{n\mathbf{k}}(\omega) \) can be expressed in terms of the bare propagator \( G_{n\mathbf{k}}(\omega) \) and the self-energy  \( \Sigma_{n\mathbf{k}}(\omega) \) through the Dyson equation as,\cite{Mahan2000,Antonius2015}

\begin{equation}
	G_{n\mathbf{k}}(\omega) = \frac{1}{\left[G_{n\mathbf{k}}^0(\omega)\right]^{-1} - \Sigma_{n\mathbf{k}}(\omega)},
	\label{eq:4_retarded_green's_function}
\end{equation}
The spectral function \( A_{n\mathbf{k}}(\omega) \) is then given by:

\begin{equation}
	A_{n\mathbf{k}}(\omega) = -\frac{1}{\pi} \text{Im} G^{R}_{n\mathbf{k}}(\omega),
	\label{eq:5_DM_spectral_function}
\end{equation}

where \( G^{R}_{n\mathbf{k}}(\omega) \) is the retarded Green's function. Combining \cref{eq:4_retarded_green's_function} and \cref{eq:5_DM_spectral_function}, the spectral function can be expressed as:

\begin{equation}
	A_{n\mathbf{k}}(\omega) = -\frac{1}{\pi} \frac{\Im m \Sigma_{n\mathbf{k}}(\omega)}{(\omega - \epsilon_{n\mathbf{k}} - \Re e \Sigma_{n\mathbf{k}}(\omega))^2 + (\Im m \Sigma_{n\mathbf{k}}(\omega))^2},
	\label{eq:6_full_exp_spectral_function}
\end{equation}

where \( \epsilon_{n\mathbf{k}} \) represents the energy of the bare state, and \( \text{Re} \Sigma_{n\mathbf{k}}(\omega) \) and \( \text{Im} \Sigma_{n\mathbf{k}}(\omega) \) are the real and imaginary parts of the self-energy, respectively.

\subsubsection{Cumulant-expansion framework}

The CE method provides a higher-order approximation to the Green's function, capturing more complex e-ph interactions. The Green's function in the time domain can be expressed as\cite{Kubo1962}:

\begin{equation}
	G_{n\mathbf{k}}(t) = G_{n\mathbf{k}}^0(t) e^{C_{n\mathbf{k}}(t)}
	\label{eq:7_cumulant_greens_function}
\end{equation}

where \( C_{n\mathbf{k}}(t) \) is the cumulant function. For e-ph interactions, the second-order cumulant is often sufficient and is exact for a fully localized core electron interacting with a phonon.\cite{gumhalter2016cumulant,guzzo2012} The cumulant functions can be derived by expanding the exponential in \cref{eq:7_cumulant_greens_function} and matching the resulting terms with the standard Feynman expansion of the Green's function.\cite{aryasetiawan1996} Utilizing the FM self-energy, which includes both Fan and Debye-Waller terms, results in the following expression,

\begin{equation}
	C_{n\mathbf{k}}(t) = \int_{-\infty}^{\infty} d\omega \beta_{n\mathbf{k}}(\omega) \frac{e^{-i\omega t} + i\omega t - 1}{\omega^2}.
	\label{eq:8_cumulant_expansion}
\end{equation}
The three terms above impact the spectral function in distinct ways. The first term gives rise to satellites, the second term shifts the QP peak, and the third term adjusts the QP weight. In \cref{eq:8_cumulant_expansion}, $\beta_{n\mathbf{k}}(\omega)$ can be obtained from the Fan self-energy as,
\begin{equation}
	\beta_{n\mathbf{k}}(\omega) = \frac{1}{\pi} \left|\text{Im} \Sigma^{Fan}_{n\mathbf{k}}(\omega + \epsilon_{n\mathbf{k}})\right| .
	\label{eq:9_boson_spectrum}
\end{equation}
The static Debye-Waller (DW) self-energy manifests as a pure shift of the QP energy as,
\begin{equation}
G_{n\mathbf{k}}(t) = -i \theta(t) e^{-i (\epsilon_{n\mathbf{k}} + \Sigma_{n\mathbf{k}}^{\text{DW}}) t} e^{C_{n\mathbf{k}}(t)}.
\label{eq:10_final_cumulant_greens_function}
\end{equation}
The spectral function is derived by applying a Fourier transform to \cref{eq:10_final_cumulant_greens_function} in the time domain, and then substituting the result into \cref{eq:5_DM_spectral_function}. Within CE framework, the Green’s function is formally exact up to second order in the e-ph interaction, while higher-order contributions are incorporated approximately. It accounts for multiple phonon satellites and provides a more accurate depiction of the spectral features. In contrast, the DM approach retains only the exact second-order terms.\cite{Hedin1980,Guzzo2011} At long times, the cumulant exhibits an affine asymptotic form, encoding both the quasiparticle lifetime and the spectral lineshape,  
\begin{equation}
G_{n\mathbf{k}}(t \to \infty) \approx G_{n\mathbf{k}}^{0}(t) \,
\exp\left[ -w_{n\mathbf{k}} + \left(-\Gamma_{n\mathbf{k}} + i\Lambda_{n\mathbf{k}}\right) t \right],
\label{eq:11_cumulant_t_inf}
\end{equation}
where \( w_{n\mathbf{k}} \) is a constant offset,\cite{Gumhalter2005} \( \Gamma_{n\mathbf{k}} = \left| \Im m\Sigma_{n\mathbf{k}}(\omega = \epsilon_{n\mathbf{k}}^{0}) \right| \) represents the decay rate, and \( \Lambda_{n\mathbf{k}} = \Re e \Sigma_{n\mathbf{k}}(\omega = \epsilon_{n\mathbf{k}}^{0}) \) denotes the energy shift. Here, \( G^{0} \) depends on \( t \) through the exponential of the bare electron energy \(\epsilon_{n\mathbf{k}}^{0}\).

\textcolor{black}{In addition to the standard DM and CE formalisms, self-consistent Green's function approaches-such as the $\text{sc}G\Delta_{0}$ scheme recently examined by Jae-Mo \textit{et al.}\cite{JMo2025} have been developed to iteratively update the self-energy and Green's function until convergence. For materials in the weak e-ph coupling regime, including Si at $T = 300 K$, they reported that the spectral functions from $\text{sc}G\Delta_{0}$ are nearly identical to those from CE, both in QP dispersion and in the fine details of the lineshape. This close agreement indicates that self-consistency has minimal impact in such systems. In the present work, we therefore adopt the CE framework for Si and Ge without further self-consistent corrections. We note that in systems with stronger e-ph coupling, for example in polar oxides, self-consistent schemes may produce more substantial modifications to the spectral features.
}

\section{Electron Mobility}

In this section, we explore the computation of electron mobility and conductivity through the Kubo-Greenwood formalism. We incorporate spectral functions derived from both the DM approach and the CE method. Notably, we have implemented the DM+ Kubo-Greenwood approach within ABINIT for this study, to facilitate comparison.

The static conductivity \( \sigma(T) \) at temperature \( T \)  is calculated using the Kubo-Greenwood formalism as,\cite{Mahan2000}

\begin{equation}
\begin{split}
    \sigma(T) = \frac{\pi e^2}{\Omega} \sum_{mn\mathbf{k}} \Re e \left( \mathbf{v}_{mn \mathbf{k}} \otimes \mathbf{v}_{nm \mathbf{k}} \right)\\ \ \  \int A_{n \mathbf{k}} (T, \omega' ) A_{m \mathbf{k}} (T,\omega') \left( - \frac{\partial f(T)}{\partial \omega'} \right) d \omega',
    \end{split}
    \label{eq:12_Kubo_greenwood_conductivity}
\end{equation}

where \( \Omega \) is the volume of the unit cell, \( e \) is the elementary charge, \( v_{nk} \) and \( v_{n'k} \) are the band velocities of the electronic states \( \vert n\mathbf{k} \rvert \) and \( \vert m\mathbf{k} \rvert \), respectively. The term \( A_{n\mathbf{k}}(T, \omega') \) is the spectral function at temperature \( T \), and \( f(T) \) is the Fermi-Dirac distribution at temperature \( T \). The term \( \Re e \left(\mathbf{v}_{n\mathbf{k}} \otimes \mathbf{v}_{m\mathbf{k}}\right) \) represents the real part of the outer product of the band velocities, capturing the directional dependence of the conductivity. The integration over \( \omega' \) accounts for the excitation of electrons due to phonons, influencing the spectral function and electron mobility at temperature \( T \). The microscopic effects of e-ph scattering are encapsulated in the spectral function. The Kubo formula is often applied in its diagonal approximation where the $n \neq m$ off-diagonal components of the velocity operator are ignored. This approximation is generally adequate when considering transport bands which are isolated at the Fermi level (e.g. simple metals).

The electron mobility, \( \mu \), is then determined using the relation,
\begin{equation}
	 \mu = \sigma / (n_c e),
     \label{eq:13_mobility}
\end{equation} 
where \( n_c \) is the carrier concentration given by,
\begin{equation}
n_c = \sum_{n\mathbf{k}} \int_{-\infty}^{\infty} d\omega \, f(\omega) A_{n\mathbf{k}}(T, \omega).
\label{eq:14_carrier_concentration}
\end{equation}
In practice we will fix $n_c$ and adjust the chemical potential.

Combining the Kubo-Greenwood formalism with the cumulant-derived spectral function enables a more precise calculation of electron mobility, especially in materials where e-ph interactions significantly influence transport properties. \textcolor{black}{This approach has proven essential in systems like SrTiO$_3$, where Fröhlich-type polar coupling dominates transport behavior,\cite{Zhou2019} and in TiO$_2$,\cite{Verdi2017} where non-adiabatic corrections and satellite features play a key role. Although silicon and germanium exhibit relatively weak electron–phonon coupling and do not conform to the standard Fröhlich picture, their transport properties still benefit from a many-body treatment. The cumulant formalism captures subtle but important spectral features—such as asymmetric broadening and incoherent satellite weight—that are not accessible through simpler approximations. These effects, particularly near band edges, can influence carrier dynamics in ways that may not be evident from mobility values alone. Furthermore, the cumulant approach provides a rigorous non-perturbative benchmark, enabling a critical assessment of widely used approximations and offering deeper insight into the underlying spectral mechanisms governing charge transport.}

Mobility calculations were performed for silicon and germanium in both p-doped and n-doped regimes, with electron and hole concentrations ranging from \(10^{17}\) to \(10^{20}\) cm\(^{-3}\). \textcolor{black}{To account for ionized impurity scattering, we applied corrections using the Brooks–Herring model.\cite{Brooks1951} This scattering mechanism becomes increasingly relevant at higher doping levels and helps bridge the gap between theory and experiment. Experimental samples contain  charged impurities and defects, 
which contribute to enhanced scattering beyond the effect of nominal dopant concentration with rigid band doping. Including this effect improves agreement with measured mobilities, particularly in the heavily doped regime. By including the impurity-limited mobility $\mu_{I}$ and combining it with the phonon-limited component $\mu_{e-ph}$ via Matthiessen’s rule,
\begin{equation}
\frac{1}{\mu_{\text{total}}} = \frac{1}{\mu_{\text{e-ph}}} + \frac{1}{\mu_{\text{imp}}},
\label{eq:15_total_scattering_rate}
\end{equation}
we obtain total mobilities that closely match experimental observations across a wide range of doping concentrations and temperatures. In the above equation, \( \mu_{\text{e-ph}} \) is the phonon-limited mobility (from CE), and \( \mu_{\text{imp}} \) is estimated from the dopant concentration and relevant screening parameters.
}

	\section{RESULTS AND DISCUSSION}

This study utilizes the ABINIT software\cite{Verstraete2025,Gonze2020,Romero2020} package to perform all calculations. Analysis of spectral functions for both doped and undoped silicon and germanium are presented in this section followed by mobility and conductivity across various doping levels. The calculations were conducted using matrix elements constructed from the self consistent DFPT potential, interpolated by Fourier transform as in \cite{Brunin2020, Brunin2020prb,Eiguren2008}. Further information on convergence tests, ground-state and DFPT computations, including parameters such as lattice constants and band gaps, is provided in \cref{appendix:computational_detail,appendix:convergence study}.

\subsection{Spectral function at finite temperature}
Once the self-energy has been converged as shown in Appendix, the spectral function can be computed at finite temperatures using both the DM and CE methods. This analysis is performed for intrinsic as well as for \( p \)-type and \( n \)-type doped silicon and germanium.

\subsubsection{Undoped Silicon and Germanium}
\begin{figure*}[t]
	\centering
	\includegraphics[width = 19.5 cm]{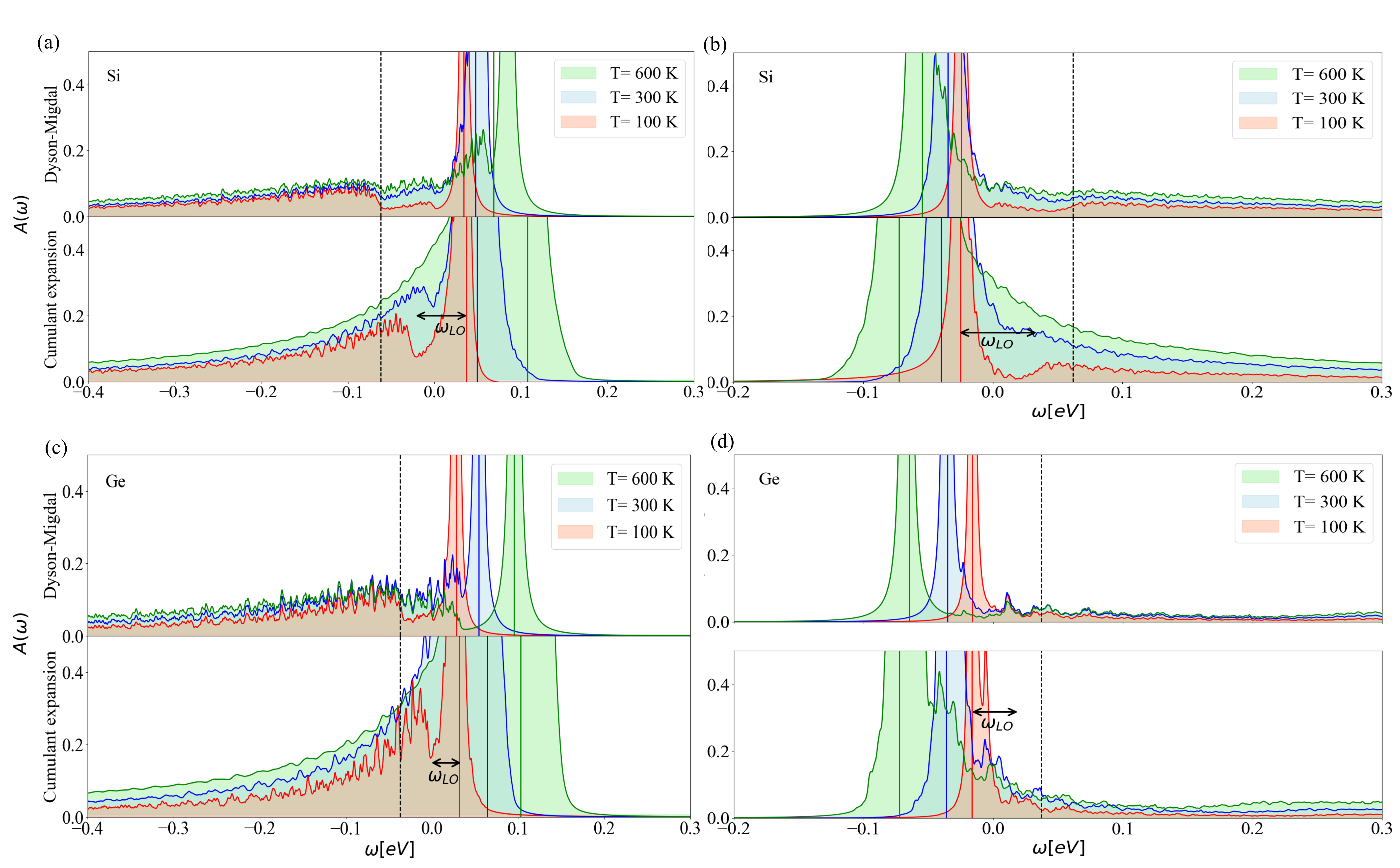}
	\caption{Spectral functions for silicon (Si) and germanium (Ge) are shown in panels (a)–(d), computed at the CBM and VBM for temperatures of 100 K, 300 K, and 600 K using both the DM and CE methods. The vertical colored lines denote the QP energies obtained from the NL approach, while the vertical dashed black lines mark the longitudinal optical phonon frequency ($\omega_{LO}$), relative to the Kohn-Sham (KS) energy:  the CE spectral function has a polaron feature $\omega_{LO}$ above the QP peak, while the DM feature is erroneously $\omega_{LO}$ above the KS energy. The broadening parameter \(\eta\) is set to 1 meV for both Si and Ge. The Brillouin zone sampling uses \(N = 160\) for Si and \(N = 120\) for Ge.}
	\label{fig:undoped}
\end{figure*}

Figures \ref{fig:undoped} (a–d) display the spectral functions computed at the VBM and CBM for silicon and germanium at temperatures \(T = 100\), \(300\), and \(600 \, \text{K}\), using both the DM and CE approaches. Across both materials, a temperature-dependent shift and broadening of the QP peak are evident. The QP energy aligns with the solution of the nonlinear QP equation (DM-NL), while the Kohn-Sham (KS) energy is fixed at \(\omega = 0\). \textcolor{black}{Compared to diamond,\cite{verstraete2024polaronic} where well-separated satellite structures form a broad spectral plateau, the behavior in Si and Ge is more nuanced. Owing to their relatively low LO phonon energies, the satellite features lie closer to the QP peak. In the CBM of both materials, distinct satellite features are still resolved, while in the VBM—particularly for germanium, the satellite partially overlaps with the QP peak, leading to a less clearly separated sideband structure.}

At \( T = 100 \, \text{K} \), the DM formalism predicts an energy separation between the QP peak and the first satellite at the VBM of approximately \( 1.5 \, \omega_{\text{LO}} \) for silicon and \( 1.7 \, \omega_{\text{LO}} \) for germanium, as shown in Figures~\ref{fig:undoped}(a) and (c). These values represent an overestimation relative to physically expected polaronic shifts. The QP energy shift with respect to the KS reference at both the CBM and VBM leads to a temperature-dependent narrowing of the band gap, as illustrated in Figures \ref{fig:undoped}(a–d). In silicon, pronounced satellite features appear in both the CBM and VBM spectral functions, indicating comparable e-ph coupling strengths at both band edges [Figures~\ref{fig:undoped}(a) and (b)]. In contrast, germanium exhibits an asymmetric behavior. While the VBM spectrum [Figure~\ref{fig:undoped}(c)] displays clear sideband structures similar to those in silicon, the CBM spectral function [Figure~\ref{fig:undoped}(d)] shows significantly weaker satellite features and minimal broadening, particularly within the CE approach. This disparity suggests that e-ph interactions are more pronounced at the VBM than at the CBM in germanium—a contrast to the more symmetric coupling observed in silicon. This asymmetry may originate from differences in the electronic structure of Ge near the band edges, such as variations in orbital composition or density of states. The suppressed phonon coupling at the germanium CBM reduces the extent of spectral weight redistribution and limits the emergence of satellite features, especially at low to intermediate temperatures. \textcolor{black}{Accordingly, the weaker CBM satellite intensity and narrower QP linewidth in the CE analysis indicate a smaller on-shell imaginary self-energy and longer conduction-electron lifetimes, consistent with Ge’s higher electron mobility. While such spectral traits can serve as qualitative indicators for identifying potential high-mobility materials - favoring cases with narrow linewidths, large QP weights, and minimal satellite intensity at the transport band edge - reliable mobility predictions still require full transport calculations.}

As the temperature increases,  the QP energies obtained from the DM nonlinear (DM-NL) approach deviate substantially from those predicted by the CE method for both silicon and germanium. This divergence underscores the improved accuracy of the CE formalism, which incorporates higher-order e-ph interactions and yields more reliable estimates of both QP energies and satellite positions. With increasing temperature, additional states below the KS energy become accessible, driven by contributions from the self-energy \(\Sigma^{-}\). The nonpolaron satellites do not move appreciably with T, an artifact of the DM approach. At high temperatures, contributions from \(\Sigma^+\) start at \(\omega_{\text{LO}}\) above the KS energy, irrespective of the QP peak position.
However, this does not reflect an actual energy shift with respect to the QP; rather, it indicates the presence of an independent satellite peak.

Within the CE framework, the spectral function at the CBM reveals that at low temperatures, the self-energy component \(\Sigma^+\) gives rise to a satellite feature located approximately \(+\omega_{\text{LO}}\) above the QP peak. As temperature increases, the QP peak broadens due to enhanced e-ph scattering, making it increasingly difficult to resolve distinct satellite features in the spectrum. Unlike the DM approach, where the self-energy is explicitly separated into \(\Sigma^+\) and \(\Sigma^-\) terms, the CE method encodes these interactions within an exponential formulation. This treatment naturally incorporates higher-order processes, leading to a smoother spectral shape and reduced separation between the QP peak and accompanying satellites at elevated temperatures. A similar temperature-dependent trend is observed at the VBM. However, unlike polar materials where the CE method often reveals distinct multiple phonon sidebands (e.g., at \(+2\omega_{\text{LO}}\)), no such well-resolved higher-order replicas are observed here. Furthermore, in contrast to diamond, where polaronic satellites emerge as a broad plateau, the satellite features in Si and Ge appear to merge with the QP peak. A detail on cumulant functions used for obtaining CE spectral functions is given in \cref{appendix:cumulant_function}.

\subsubsection{Doped Silicon}

\begin{figure}[t]
	\centering
	\includegraphics[width = 9.5 cm]{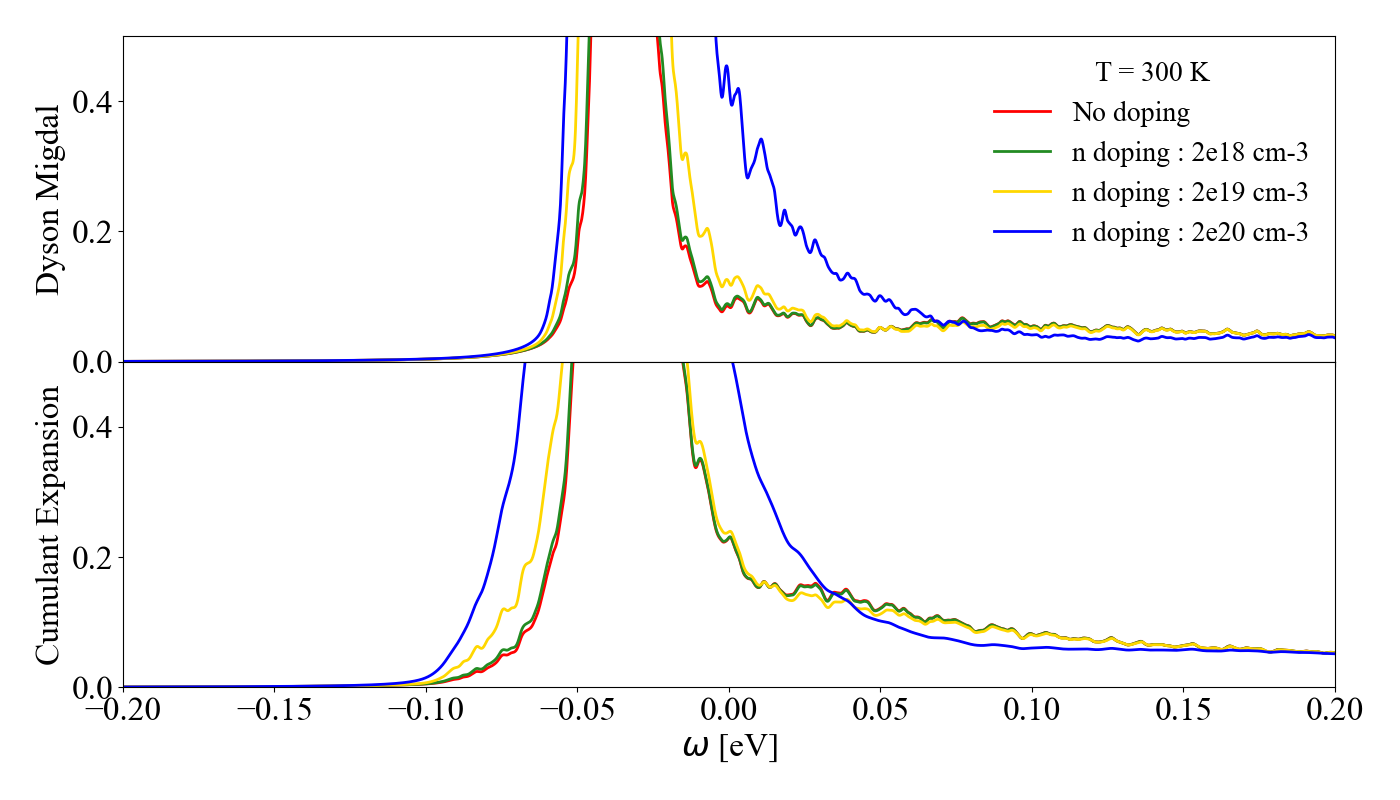}
	\caption{Silicon spectral functions calculated at the CBM for n-type doping levels $2 \times 10^{18}$, $2 \times 10^{19}$ and $2 \times 10^{20}$   using the DM and CE methods at T=300 K. The broadening parameter $\eta$ is set to 1 meV and N = 160.}
	\label{fig:dopedcbm_si}
\end{figure}

\begin{figure}[t]
	\centering
	\includegraphics[width = 9.5 cm]{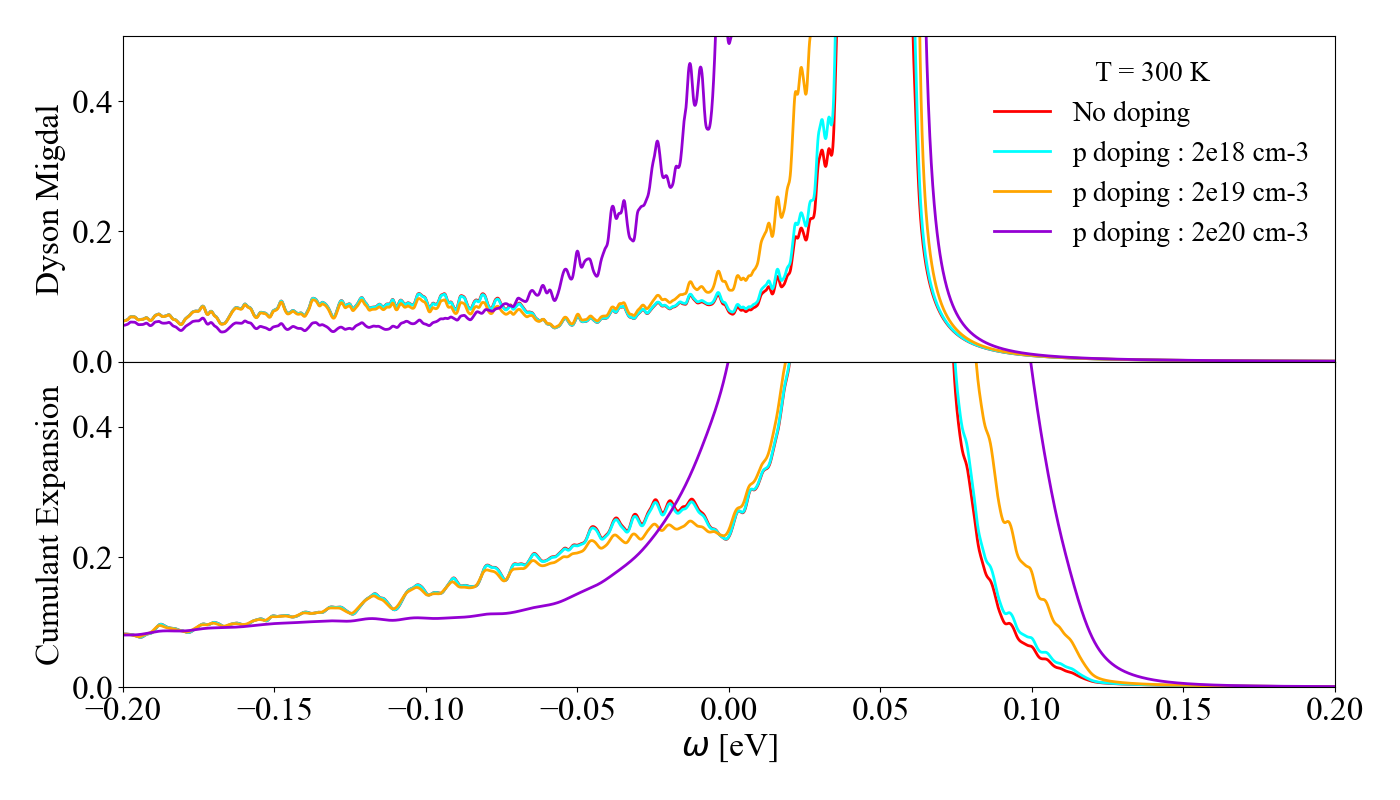}
	\caption{Silicon spectral functions calculated at the VBM for for p-type doping levels 2e18, 2e19 and 2e20   using the DM and CE methods at T=300 K. The broadening parameter $\eta$ is set to 1 meV and N = 160.}
	\label{fig:dopedvbm_si}
\end{figure}

The spectral functions for doped silicon, shown in Figures \ref{fig:dopedcbm_si} and \ref{fig:dopedvbm_si}, provide insights into the effects of \( n \)-type and \( p \)-type doping on the CBM and VBM, respectively. The results highlight how varying doping concentrations influence the QP peak and its associated satellite features at \( T = 300 \, \text{K} \).

In \( n \)-type doped silicon (Figure \ref{fig:dopedcbm_si}), increasing the electron concentration from \( 2 \times 10^{18} \, \text{cm}^{-3} \) to \( 2 \times 10^{20} \, \text{cm}^{-3} \) leads to a significant broadening of the QP peak. At higher doping levels, the CE spectral function shows a shift in the QP peak, accompanied by increased spectral weight in the tail region. This behavior is attributed to enhanced e-ph interactions due to the increased carrier concentration and the introduction of additional electronic states. Consistent with observations in undoped silicon, the DM approach inaccurately positions the satellite peaks. The broadening effect is more pronounced in the CE method, which accounts for higher-order scattering processes. Notably, \( n \)-type doping does not significantly affect the spectral functions at the VBM.

In \( p \)-type doped silicon (Figure \ref{fig:dopedvbm_si}), increasing hole concentrations result in a broadening and noticeable shift of the QP peak at the VBM. At doping levels up to \( 2 \times 10^{20} \, \text{cm}^{-3} \), the QP peak broadens considerably due to strong many-body interactions.

Relative to the undoped case, enhanced \( p \)-type and \( n \)-type doping inherently incorporates higher-order contributions, leading to smoother spectral peaks and a reduced separation between satellites and the QP peak. These findings underscore the significant role of doping in altering silicon's spectral characteristics, with \( n \)-type doping primarily influencing the CBM and \( p \)-type doping affecting the VBM. 

\textcolor{black}{Doping introduces additional free carriers that modify spectral features through two main effects: an increase in the available phase space for e-ph scattering near the Fermi level, which raises scattering rates, and changes in the screening of Coulomb interactions. In polar materials, the latter can strongly reduce long-range Fr\"ohlich coupling and reshape satellite structure as seen in TiO$_2$.\cite{Verdi2017} By contrast, Si and Ge are nonpolar with room-temperature permittivities \(\varepsilon_r^{\mathrm{Si}}\!\approx\!11.7\) and \(\varepsilon_r^{\mathrm{Ge}}\!\approx\!16.2\),\cite{Levinshtein2001-va} so free-carrier screening primarily affects long-range Coulomb fields, such as those from ionized impurities, and only weakly renormalizes the short-range potentials that dominate their phonon-limited transport. Thus, doping-dependent renormalization of the e-ph matrix elements is not included in our present simulations.
}

\begin{figure}[thb]
	\centering
	\includegraphics[width = 9.5 cm]{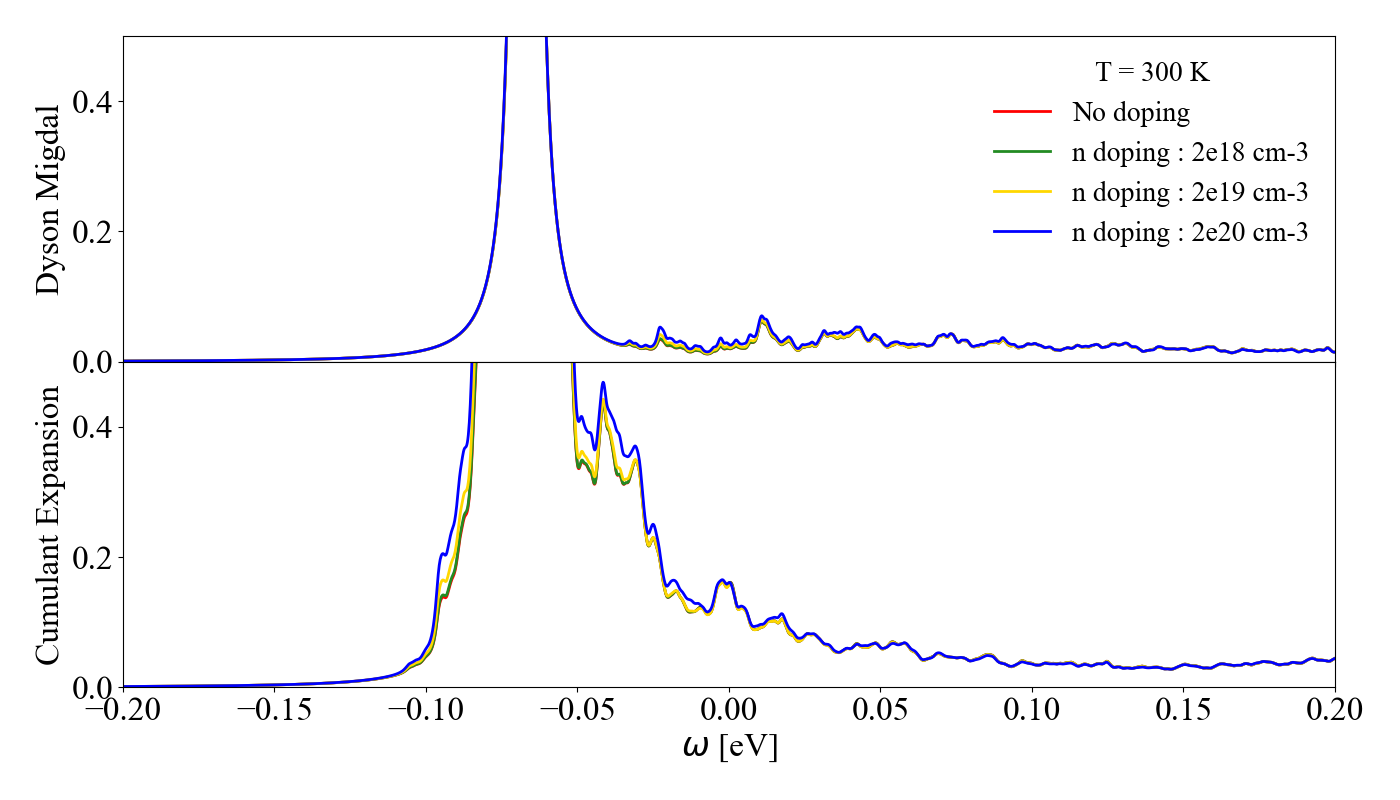}
	\caption{Germanium spectral functions calculated at the CBM for n-type doping levels $2 \times 10^{18}$, $2 \times 10^{19}$ and $2 \times 10^{20}$ cm$^{-3}$   using the DM and CE methods at T=300 K. The broadening parameter $\eta$ is set to 1 meV and N = 160.}
	\label{fig:dopedcbm_ge}
\end{figure}
\begin{figure}[thb]
	\centering
	\includegraphics[width = 9.5 cm]{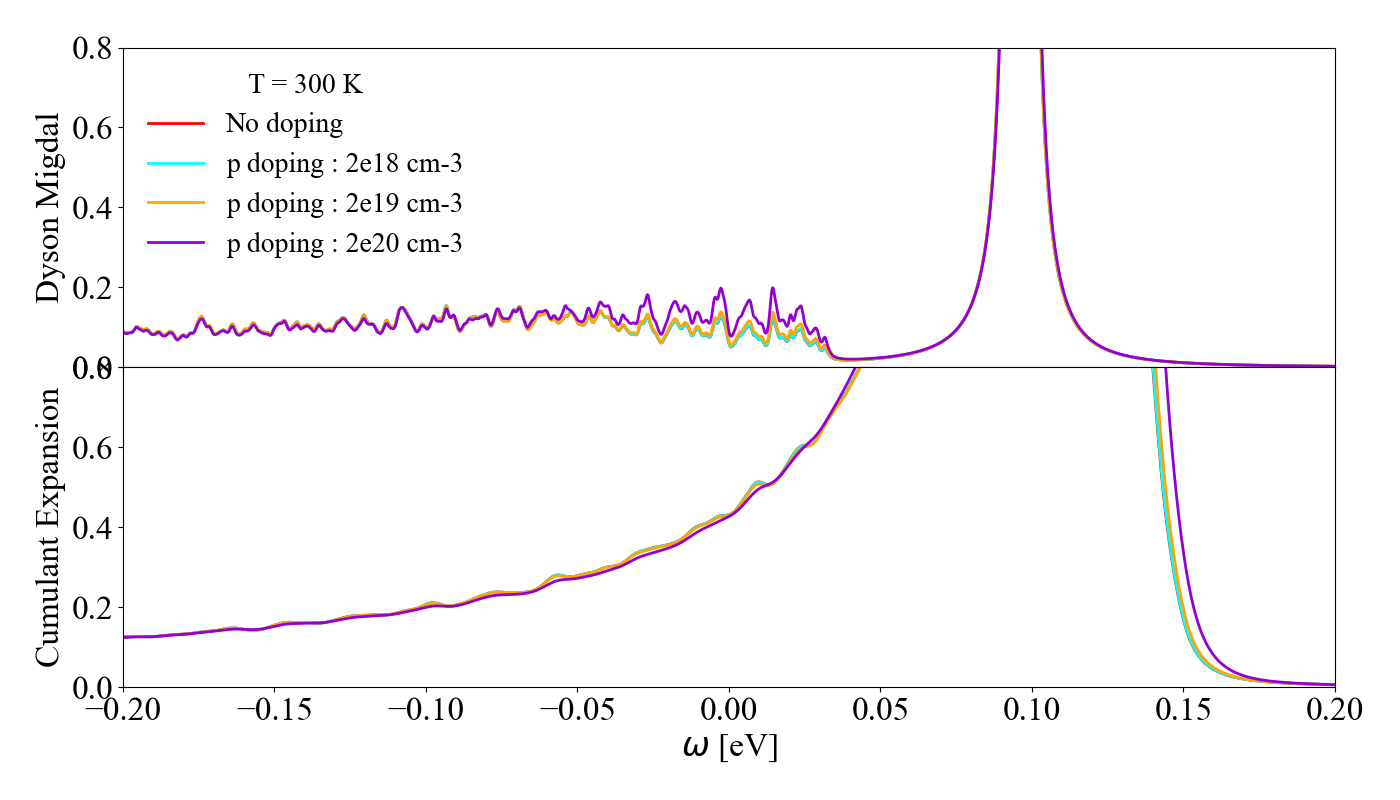}
	\caption{Germanium spectral functions calculated at the VBM for for p-type doping levels 2e18, 2e19 and 2e20 cm$^{-3}$   using the DM and CE methods at T=300 K. The broadening parameter $\eta$ is set to 1 meV and N = 160.}
	\label{fig:dopedvbm_ge}
\end{figure}
\subsubsection{Doped Germanium}

The spectral functions for doped germanium, shown in Figures~\ref{fig:dopedcbm_ge} and \ref{fig:dopedvbm_ge}, reveal the influence of $n$-type and $p$-type doping on the CBM and VBM, respectively, at $T = 300 \, \mathrm{K}$. Compared to silicon, the doping-induced modifications in germanium are qualitatively similar but quantitatively less pronounced.

In $n$-type doped germanium (Figure~\ref{fig:dopedcbm_ge}), increasing the electron concentration from $2 \times 10^{18} \, \mathrm{cm}^{-3}$ to $2 \times 10^{20} \, \mathrm{cm}^{-3}$ leads to a moderate broadening of the QP peak in the CE spectral function. The satellite features gain additional spectral weight at higher doping levels, though the overall QP peak position and broadening are less sensitive to doping compared to silicon. The DM approach continues to underestimate the changes, failing to capture the correct redistribution of spectral weight into the satellite region. The relatively weaker effect of $n$-type doping on the CBM of germanium is consistent with its smaller e-ph coupling strength and softer phonon modes compared to silicon. \textcolor{black}{This reduction in scattering strength, partly due to the lower phonon energies, contributes to longer carrier lifetimes and consequently higher electron mobility in germanium.} As expected, $n$-type doping does not noticeably alter the spectral features at the VBM.

For $p$-type doped germanium (Figure~\ref{fig:dopedvbm_ge}), increasing the hole concentration up to $2 \times 10^{20} \, \mathrm{cm}^{-3}$ results in minor broadening and a small shift of the QP peak at the VBM in the CE method. Satellite formation is weak, and the QP peak remains sharp even at high doping levels, reflecting the reduced hole-phonon coupling strength in germanium. The DM approach again shows minimal changes, unable to capture the subtle redistribution of spectral weight observed in the CE results. Unlike in silicon, where $p$-type doping causes substantial many-body broadening at the VBM, the effects in germanium remain small, likely due to its heavier valence band and weaker deformation potentials.

These results confirm that while doping enhances e-ph interactions and many-body scattering in both materials, the magnitude of these effects is very material-dependent. Germanium, with intrinsically weaker e-ph coupling, exhibits a much more modest evolution of its spectral functions with doping. As in silicon, the CE method provides a more accurate description of the spectral features at elevated carrier concentrations, highlighting the importance of higher-order processes in describing QP broadening and satellite formation in doped semiconductors.

\subsection{Role of Short-Range and Long-Range Contributions}

To better understand the spatial characteristics of e-ph coupling in silicon and germanium, we applied the Fourier-space interpolation method discussed in Ref. \cite{verstraete2024polaronic}. This approach enables us to decompose the unit-cell averaged local scattering potential into contributions from short-range (SR) crystal fields and long-range (LR) multipolar fields, including quadrupolar interactions (the dipolar terms are 0 by symmetry in covalent Si and Ge). The absolute values of these components, resolved along the Cartesian directions $\hat{x}$, $\hat{y}$, and $\hat{z}$, are plotted along high-symmetry paths in the Brillouin zone for both materials.
\begin{figure}[thb]
	\centering
	\includegraphics[width = 9.5 cm]{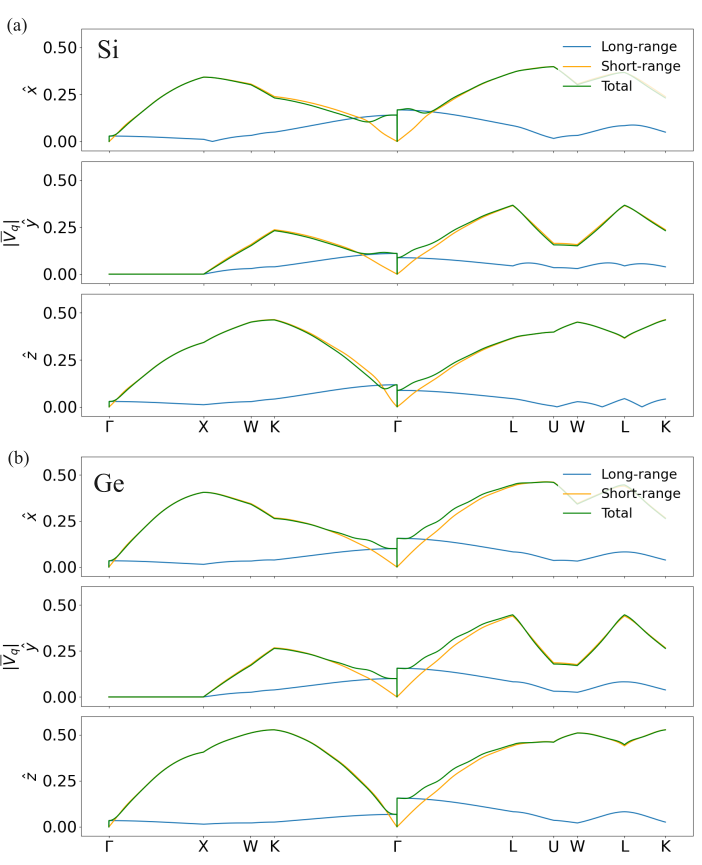}
	\caption{Unit-cell averaged absolute values of the total DFPT scattering potential (green) compared to its long-range (blue) and short-range (orange) components for (a) silicon and (b) germanium. The three subplots correspond to the electron–phonon potential generated by displacing an atom along each of the three reduced directions \(\hat{x}\), \(\hat{y}\), and \(\hat{z}\). Both atoms exhibit the same \(|V_{\mathbf{q}}|\).
}
	\label{fig:pot_si_ge}
\end{figure}

\cref{fig:pot_si_ge}(a) and (b) show that in both silicon and germanium, the short-range (SR) potential dominates throughout the Brillouin zone, reflecting that local crystal field effects primarily govern electron–phonon interactions in these covalent semiconductors. The long-range (LR) contributions, although much smaller in magnitude, exhibit the singular behavior near the \(\Gamma\) point. This singular LR component is subtracted from the total potential to yield the smooth SR part, which remains finite and well-behaved. \textcolor{black}{In both materials, a discontinuity near \(\mathbf{q} \to \Gamma\) is observed in
the LR and total potentials, originating from the long–range quadrupolar field.
As discussed in Ref.~\cite{Brunin2020}, accurate long–wavelength interpolation
of e–ph matrix elements requires proper inclusion of these terms; any residual
step or divergence at \(\Gamma\) points to missing or inaccurately evaluated LR
contributions. In Si and Ge, where short–range potentials dominate, these discontinuities have a negligible impact
on the calculated spectral functions and mobilities.
}

To quantify the crossover between SR and LR interactions, we identify the $|\mathbf{q}|$ values where the two contributions become comparable. For silicon, this occurs around $|\mathbf{q}| \approx 0.14$ in reciprocal lattice units, corresponding to an approximate real-space interaction radius of \SI{22}{\angstrom}-\SI{25}{\angstrom}, or about 8-9 unit cells. For germanium, the crossover takes place slightly earlier, at $|\mathbf{q}| \approx 0.12$, implying an even longer-range interaction radius of approximately \SI{26}{\angstrom}-\SI{28}{\angstrom}, or roughly 9-10 unit cells.

These distances suggest that both silicon and germanium exhibit medium-to-large polaronic behavior, with the effect being more extended in germanium. This is consistent with the more delocalized nature of valence electrons and weaker ionic potentials in germanium compared to silicon.

In comparison, diamond,\cite{verstraete2024polaronic} has a crossover distance of $|\mathbf{q}| \approx 0.18$ and a real-space interaction length of approximately \SI{18.3}{\angstrom} (around 7 unit cells): SR interactions in diamond decay more rapidly than in Si or Ge. \textcolor{black}{This is consistent with diamond's even more strongly covalent character
which favors shorter-range deformation-potential coupling. The remaining long-range contribution is quadrupolar and comparatively weak, so the e-ph interaction is effectively
more localized.} 

Consequently, while all three materials share similar qualitative features in their e-ph scattering potentials, the range and strength of SR and LR contributions vary in a manner that reflects their underlying bonding character/strength and lattice hardness.

We emphasize that the radius deduced from the SR-LR crossover length should not be conflated with the polaron radius. The latter requires a self-consistent, many-body treatment of e-ph coupling (see, e.g., Refs.~\cite{Sio2019,Peeters1985}).

\subsection{Carrier Mobility in Doped Si and Ge}
\begin{figure}
    \centering
    \includegraphics[width=7.5cm]{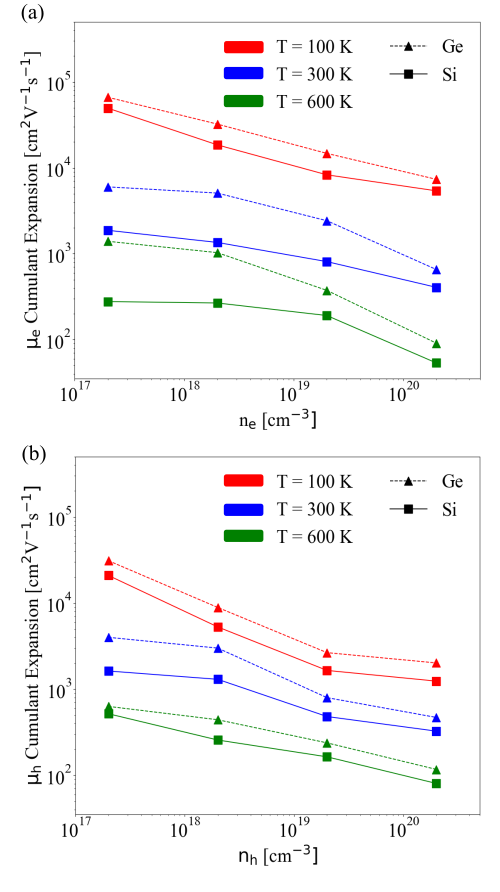}
    \caption{Variation of (a) electron mobility (\( \mu_e \)) and (b) hole mobility (\( \mu_h \)) as functions of carrier concentration and temperature, calculated using the CE spectral function. Results are shown for both Si (squares) and Ge (triangles) at \( T = 100,\,300,\,600\,\text{K} \).}
    \label{fig:mob_e_h}
\end{figure}
\begin{figure}[h]
    \centering
    \includegraphics[width=7.5cm]{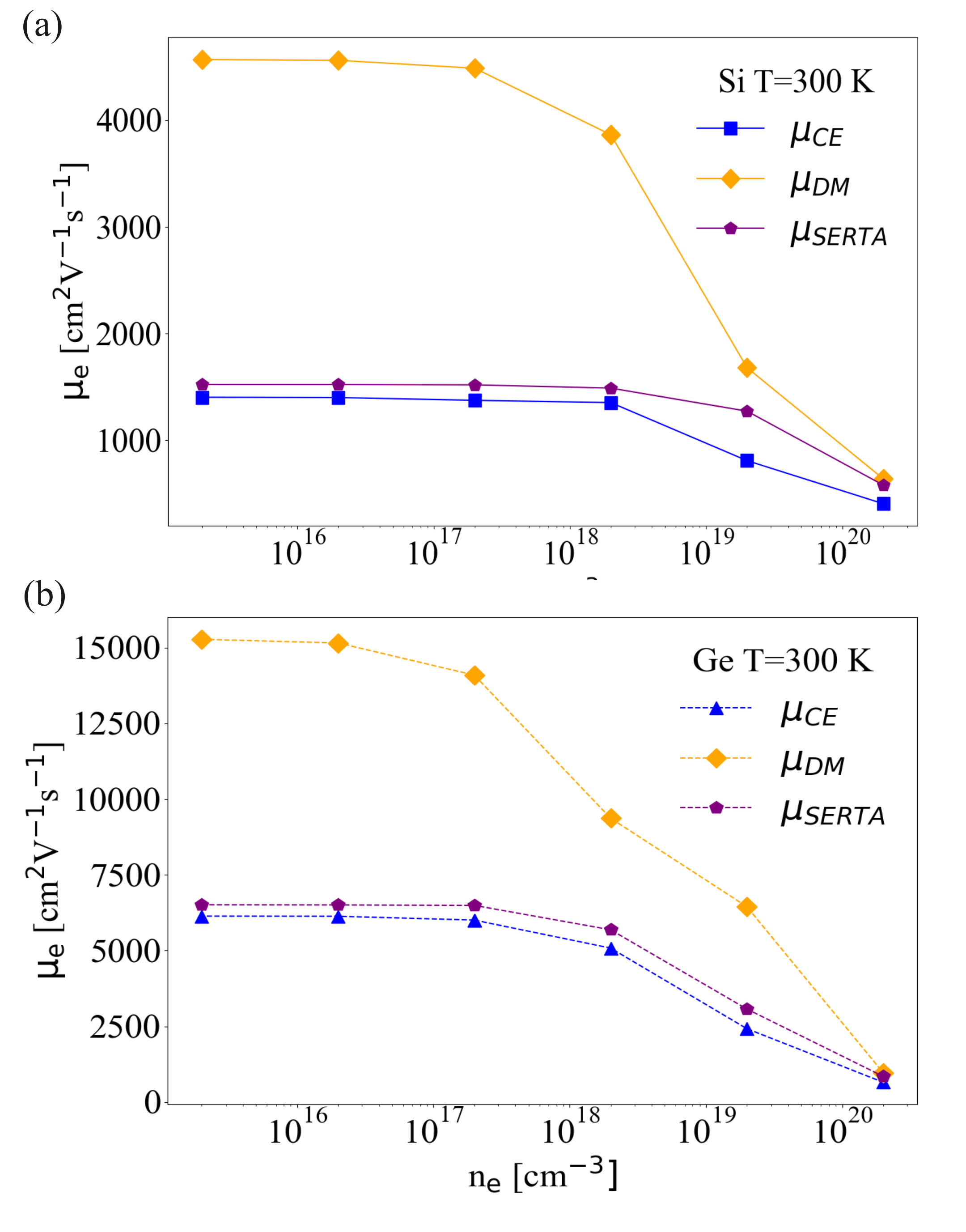}
    \caption{Comparison of electron mobility (\( \mu_e \)) at \( T = 300\,\text{K} \) in (a) Si and (b) Ge, calculated using the CE (\( \mu_{\text{CE}} \)), DM (\( \mu_{\text{DM}} \)), and SERTA (\( \mu_{\text{SERTA}} \)) approaches.}
    \label{fig:Si_Ge_SERTA}
\end{figure}

The carrier mobility calculations based on the Kubo–Greenwood formalism employ the double-grid (DG) technique to accelerate convergence while reducing computational cost, as detailed in \cref{appendix:double_grid}. In both doped silicon and germanium, carrier mobility exhibits a systematic decrease with increasing doping concentration for both \( n \)-type and \( p \)-type carriers, primarily due to enhanced scattering processes. Figures~\ref{fig:mob_e_h} and \ref{fig:Si_Ge_SERTA} present the computed electron and hole mobilities over a wide range of carrier concentrations and temperatures using different spectral treatments. For both materials, mobility decreases monotonically with doping, reflecting increased electron–phonon (e-ph) and ionized impurity scattering. Mobility also decreases with temperature due to the thermal population of phonon modes that enhance inelastic scattering.

At a given temperature and doping level, germanium consistently shows higher carrier mobility than silicon. This is due to its lower effective masses and weaker electron–phonon coupling, resulting in longer carrier lifetimes. For example, at \( T = 300\,\text{K} \), electron mobilities in Ge remain in the range of \( 5 \times 10^3 \)–\( 10^4~\mathrm{cm^2/V\cdot s} \), whereas values for Si are significantly lower across the same doping range. Hole mobilities follow a similar trend but are lower in magnitude, especially in Si, due to heavier valence-band curvature.

The calculated mobilities also depend strongly on the spectral method used. Figure~\ref{fig:Si_Ge_SERTA} compares mobility values obtained from the DM, CE, and self-energy relaxation-time approximation (SERTA) methods. The DM approach systematically overestimates mobility, particularly at low doping, because it underestimates QP broadening and ignores important vertex corrections. The CE method, which includes dynamical many-body effects beyond the lowest-order self-energy, yields more physically realistic mobilities with consistent doping and temperature trends. SERTA, which uses the imaginary part of the self-energy evaluated at the QP energy, produces results that lie between CE and DM, offering a reliable estimate of scattering-limited transport while being computationally less intensive.

While both CE and SERTA capture phonon-limited scattering with improved accuracy over DM, the CE approach also offers broader physical insight. In particular, CE resolves spectral features such as satellite peaks, and temperature-induced redistribution of spectral weight—none of which are possible in SERTA or visible in DM. These features are essential to understand not just transport, but also optical absorption, QP lifetimes, and many-body renormalization effects. The CE framework is therefore crucial when transport and spectroscopic observables are to be treated on equal theoretical footing.

\begin{figure}[h]
    \centering
    \includegraphics[width=7.5cm]{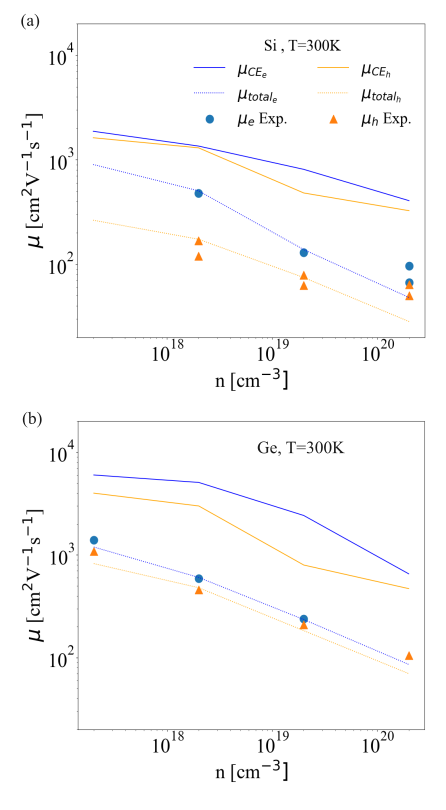}
    \caption{
Electron (\(\mu_e\), blue) and hole (\(\mu_h\), orange) mobilities at \(T = 300\,\mathrm{K}\) as a function of carrier concentration \(n\) for (a) silicon and (b) germanium. Solid lines (\(\mu_{\mathrm{CE},e}\), \(\mu_{\mathrm{CE},h}\)) denote values using the CE spectral functions. Dashed lines (\(\mu_{\mathrm{total}_e}\), \(\mu_{\mathrm{total}_h}\)) correspond to total mobilities including both e-ph and ionized impurity scattering. Experimental data\cite{canali1975electron,Jacoboni1977,Mousty1974,Norton1973,Ottaviani1975,Logan1960,Jacoboni1981,Debye1954,Spitzer1961} for electrons (blue circles) and holes (orange triangles) are shown for comparison.
}
    \label{fig:Si_Ge_exp_compare}
\end{figure}

\textcolor{black}{All mobilities reported above are only limited by phonon scattering and
therefore represent upper bounds. At higher doping, experimental samples
incur additional scattering from ionized dopants and defects that
we did not include so far. Our values exceed measurements, and the gap grows with
carrier density. This is consistent with the earlier discussion: in nonpolar
Si or Ge, free-carrier screening only weakly renormalizes short-range e-ph
couplings but does not eliminate Coulomb scattering by dopants.} To address this, we incorporate impurity-limited scattering using the Brooks–Herring model\cite{Brooks1951} and compute the total mobility as shown in \cref{fig:Si_Ge_exp_compare}. The corrected mobility shows excellent agreement with experiment across temperature and doping regimes, demonstrating the need to combine first-principles phonon scattering with empirical impurity models for predictive transport calculations in semiconductors.

\section{Conclusion}

\textcolor{black}{We have presented a comprehensive first-principles investigation of many-body effects in the spectral and transport properties of silicon and germanium, two archetypal covalent semiconductors. Our results reveal distinct nonpolaronic spectral features that require the full cumulant expansion for accurate resolution, with material-dependent amplitudes and consequences for measurable properties. Temperature-dependent spectral functions show that Si exhibits relatively symmetric electron–phonon coupling at both band edges, while Ge displays pronounced asymmetry, with stronger coupling at the valence band maximum and much weaker coupling at the conduction band minimum. This asymmetry correlates with Ge’s consistently higher electron mobility, as reduced CBM coupling leads to longer carrier lifetimes and lower phonon-limited scattering rates.
Doping studies demonstrate that increased carrier concentrations broaden spectral peaks and compress QP–satellite separation by increasing the available phase space for scattering near the Fermi level. These effects are more pronounced in Si than in Ge, reflecting Ge’s intrinsically weaker electron–phonon coupling. The DFPT scattering potentials confirm that short-range crystal-field interactions dominate in both materials, consistent with their nonpolar, covalent bonding character. 
Mobility calculations performed within the Kubo–Greenwood formalism, and supplemented by ionized-impurity scattering models, reproduce experimental trends across doping levels. The trends in mobility are directly linked to the spectral features: broader QP peaks and enhanced satellite weight correspond to stronger scattering and reduced mobility. The agreement with experiment highlights the importance of combining first-principles phonon-limited transport with realistic defect-scattering models when modeling doped semiconductors.
Overall, this work establishes the cumulant expansion as an essential tool for describing electron–phonon coupling in nonpolar semiconductors, providing quantitative links between spectroscopic fingerprints and transport properties. The methodology and insights developed here are broadly applicable to other covalent systems, enabling predictive modeling of carrier dynamics in technologically relevant materials.
}

\bibliography{reference}

\begin{thebibliography}{63}%
\makeatletter
\providecommand \@ifxundefined [1]{%
 \@ifx{#1\undefined}
}%
\providecommand \@ifnum [1]{%
 \ifnum #1\expandafter \@firstoftwo
 \else \expandafter \@secondoftwo
 \fi
}%
\providecommand \@ifx [1]{%
 \ifx #1\expandafter \@firstoftwo
 \else \expandafter \@secondoftwo
 \fi
}%
\providecommand \natexlab [1]{#1}%
\providecommand \enquote  [1]{``#1''}%
\providecommand \bibnamefont  [1]{#1}%
\providecommand \bibfnamefont [1]{#1}%
\providecommand \citenamefont [1]{#1}%
\providecommand \href@noop [0]{\@secondoftwo}%
\providecommand \href [0]{\begingroup \@sanitize@url \@href}%
\providecommand \@href[1]{\@@startlink{#1}\@@href}%
\providecommand \@@href[1]{\endgroup#1\@@endlink}%
\providecommand \@sanitize@url [0]{\catcode `\\12\catcode `\$12\catcode
  `\&12\catcode `\#12\catcode `\^12\catcode `\_12\catcode `\%12\relax}%
\providecommand \@@startlink[1]{}%
\providecommand \@@endlink[0]{}%
\providecommand \url  [0]{\begingroup\@sanitize@url \@url }%
\providecommand \@url [1]{\endgroup\@href {#1}{\urlprefix }}%
\providecommand \urlprefix  [0]{URL }%
\providecommand \Eprint [0]{\href }%
\providecommand \doibase [0]{https://doi.org/}%
\providecommand \selectlanguage [0]{\@gobble}%
\providecommand \bibinfo  [0]{\@secondoftwo}%
\providecommand \bibfield  [0]{\@secondoftwo}%
\providecommand \translation [1]{[#1]}%
\providecommand \BibitemOpen [0]{}%
\providecommand \bibitemStop [0]{}%
\providecommand \bibitemNoStop [0]{.\EOS\space}%
\providecommand \EOS [0]{\spacefactor3000\relax}%
\providecommand \BibitemShut  [1]{\csname bibitem#1\endcsname}%
\let\auto@bib@innerbib\@empty
\bibitem [{\citenamefont {Jacoboni}\ and\ \citenamefont
  {Lugli}(1989)}]{jacoboni1989monte}%
  \BibitemOpen
  \bibfield  {author} {\bibinfo {author} {\bibfnamefont {C.}~\bibnamefont
  {Jacoboni}}\ and\ \bibinfo {author} {\bibfnamefont {P.}~\bibnamefont
  {Lugli}},\ }\href@noop {} {\emph {\bibinfo {title} {The Monte Carlo Method
  for Semiconductor Device Simulation}}}\ (\bibinfo  {publisher} {Springer},\
  \bibinfo {year} {1989})\BibitemShut {NoStop}%
\bibitem [{\citenamefont {Rode}(1975)}]{rode1975low}%
  \BibitemOpen
  \bibfield  {author} {\bibinfo {author} {\bibfnamefont {D.~L.}\ \bibnamefont
  {Rode}},\ }\bibinfo {title} {Low-field electron transport},\ in\ \href@noop
  {} {\emph {\bibinfo {booktitle} {Semiconductors and Semimetals}}},\
  Vol.~\bibinfo {volume} {10}\ (\bibinfo  {publisher} {Academic Press},\
  \bibinfo {year} {1975})\ pp.\ \bibinfo {pages} {1--89}\BibitemShut {NoStop}%
\bibitem [{\citenamefont {Canali}\ \emph {et~al.}(1975)\citenamefont {Canali}
  \emph {et~al.}}]{canali1975electron}%
  \BibitemOpen
  \bibfield  {author} {\bibinfo {author} {\bibfnamefont {C.}~\bibnamefont
  {Canali}} \emph {et~al.},\ }\bibfield  {title} {\bibinfo {title} {Electron
  and hole drift velocity measurements in silicon and their empirical relation
  to electric field and temperature},\ }\href@noop {} {\bibfield  {journal}
  {\bibinfo  {journal} {IEEE Transactions on Electron Devices}\ }\textbf
  {\bibinfo {volume} {22}},\ \bibinfo {pages} {1045} (\bibinfo {year}
  {1975})}\BibitemShut {NoStop}%
\bibitem [{\citenamefont {Debye}\ and\ \citenamefont
  {Conwell}(1952)}]{gerDebye1952}%
  \BibitemOpen
  \bibfield  {author} {\bibinfo {author} {\bibfnamefont {P.~P.}\ \bibnamefont
  {Debye}}\ and\ \bibinfo {author} {\bibfnamefont {E.~M.}\ \bibnamefont
  {Conwell}},\ }\bibfield  {title} {\bibinfo {title} {Mobility of electrons in
  germanium},\ }\href {https://doi.org/10.1103/physrev.87.1131.2} {\bibfield
  {journal} {\bibinfo  {journal} {Physical Review}\ }\textbf {\bibinfo {volume}
  {87}},\ \bibinfo {pages} {1131–1132} (\bibinfo {year} {1952})}\BibitemShut
  {NoStop}%
\bibitem [{\citenamefont {Johnson}\ and\ \citenamefont
  {Lark-Horovitz}(1950)}]{gerJohnson1950}%
  \BibitemOpen
  \bibfield  {author} {\bibinfo {author} {\bibfnamefont {V.~A.}\ \bibnamefont
  {Johnson}}\ and\ \bibinfo {author} {\bibfnamefont {K.}~\bibnamefont
  {Lark-Horovitz}},\ }\bibfield  {title} {\bibinfo {title} {Electronic mobility
  in germanium},\ }\href {https://doi.org/10.1103/physrev.79.409.2} {\bibfield
  {journal} {\bibinfo  {journal} {Physical Review}\ }\textbf {\bibinfo {volume}
  {79}},\ \bibinfo {pages} {409–410} (\bibinfo {year} {1950})}\BibitemShut
  {NoStop}%
\bibitem [{\citenamefont {Koenig}\ and\ \citenamefont
  {Gunther-Mohr}(1957)}]{gerKoenig1957}%
  \BibitemOpen
  \bibfield  {author} {\bibinfo {author} {\bibfnamefont {S.}~\bibnamefont
  {Koenig}}\ and\ \bibinfo {author} {\bibfnamefont {G.}~\bibnamefont
  {Gunther-Mohr}},\ }\bibfield  {title} {\bibinfo {title} {The low temperature
  electrical conductivity of n-type germanium},\ }\href
  {https://doi.org/10.1016/0022-3697(57)90071-9} {\bibfield  {journal}
  {\bibinfo  {journal} {Journal of Physics and Chemistry of Solids}\ }\textbf
  {\bibinfo {volume} {2}},\ \bibinfo {pages} {268–283} (\bibinfo {year}
  {1957})}\BibitemShut {NoStop}%
\bibitem [{\citenamefont {Koenig}\ \emph {et~al.}(1962)\citenamefont {Koenig},
  \citenamefont {Brown},\ and\ \citenamefont {Schillinger}}]{gerKoenig1962}%
  \BibitemOpen
  \bibfield  {author} {\bibinfo {author} {\bibfnamefont {S.~H.}\ \bibnamefont
  {Koenig}}, \bibinfo {author} {\bibfnamefont {R.~D.}\ \bibnamefont {Brown}},\
  and\ \bibinfo {author} {\bibfnamefont {W.}~\bibnamefont {Schillinger}},\
  }\bibfield  {title} {\bibinfo {title} {Electrical conduction in n-type
  germanium at low temperatures},\ }\href
  {https://doi.org/10.1103/physrev.128.1668} {\bibfield  {journal} {\bibinfo
  {journal} {Physical Review}\ }\textbf {\bibinfo {volume} {128}},\ \bibinfo
  {pages} {1668–1696} (\bibinfo {year} {1962})}\BibitemShut {NoStop}%
\bibitem [{\citenamefont {Morin}\ and\ \citenamefont
  {Maita}(1954)}]{gerMorin1954}%
  \BibitemOpen
  \bibfield  {author} {\bibinfo {author} {\bibfnamefont {F.~J.}\ \bibnamefont
  {Morin}}\ and\ \bibinfo {author} {\bibfnamefont {J.~P.}\ \bibnamefont
  {Maita}},\ }\bibfield  {title} {\bibinfo {title} {Conductivity and hall
  effect in the intrinsic range of germanium},\ }\href
  {https://doi.org/10.1103/physrev.94.1525} {\bibfield  {journal} {\bibinfo
  {journal} {Physical Review}\ }\textbf {\bibinfo {volume} {94}},\ \bibinfo
  {pages} {1525–1529} (\bibinfo {year} {1954})}\BibitemShut {NoStop}%
\bibitem [{\citenamefont {Pop}\ \emph {et~al.}(2010)\citenamefont {Pop},
  \citenamefont {Sinha},\ and\ \citenamefont {Goodson}}]{pop2010heat}%
  \BibitemOpen
  \bibfield  {author} {\bibinfo {author} {\bibfnamefont {E.}~\bibnamefont
  {Pop}}, \bibinfo {author} {\bibfnamefont {S.}~\bibnamefont {Sinha}},\ and\
  \bibinfo {author} {\bibfnamefont {K.~E.}\ \bibnamefont {Goodson}},\
  }\bibfield  {title} {\bibinfo {title} {Heat generation and transport in
  nanometer-scale transistors},\ }\href@noop {} {\bibfield  {journal} {\bibinfo
   {journal} {Proceedings of the IEEE}\ }\textbf {\bibinfo {volume} {94}},\
  \bibinfo {pages} {1587} (\bibinfo {year} {2010})}\BibitemShut {NoStop}%
\bibitem [{\citenamefont {Nayfeh}\ \emph {et~al.}(2005)\citenamefont {Nayfeh}
  \emph {et~al.}}]{nayfeh2005effects}%
  \BibitemOpen
  \bibfield  {author} {\bibinfo {author} {\bibfnamefont {O.~M.}\ \bibnamefont
  {Nayfeh}} \emph {et~al.},\ }\bibfield  {title} {\bibinfo {title} {Effects of
  high fields on electron transport in nanoscale silicon devices},\ }\href@noop
  {} {\bibfield  {journal} {\bibinfo  {journal} {IEEE Transactions on Electron
  Devices}\ }\textbf {\bibinfo {volume} {52}},\ \bibinfo {pages} {2003}
  (\bibinfo {year} {2005})}\BibitemShut {NoStop}%
\bibitem [{\citenamefont {Frohlich}(1954)}]{frohlich1954electrons}%
  \BibitemOpen
  \bibfield  {author} {\bibinfo {author} {\bibfnamefont {H.}~\bibnamefont
  {Frohlich}},\ }\bibfield  {title} {\bibinfo {title} {Electrons in lattice
  fields},\ }\href@noop {} {\bibfield  {journal} {\bibinfo  {journal} {Advances
  in Physics}\ }\textbf {\bibinfo {volume} {3}},\ \bibinfo {pages} {325}
  (\bibinfo {year} {1954})}\BibitemShut {NoStop}%
\bibitem [{\citenamefont {Emin}(1989)}]{emin1989polaron}%
  \BibitemOpen
  \bibfield  {author} {\bibinfo {author} {\bibfnamefont {D.}~\bibnamefont
  {Emin}},\ }\bibfield  {title} {\bibinfo {title} {Polaron mechanisms for
  multiphonon-assisted hopping of small polarons},\ }\href@noop {} {\bibfield
  {journal} {\bibinfo  {journal} {Solid State Communications}\ }\textbf
  {\bibinfo {volume} {70}},\ \bibinfo {pages} {127} (\bibinfo {year}
  {1989})}\BibitemShut {NoStop}%
\bibitem [{\citenamefont {Abreu}\ \emph {et~al.}(2022)\citenamefont {Abreu},
  \citenamefont {Nery}, \citenamefont {Giantomassi}, \citenamefont {Gonze},\
  and\ \citenamefont {Verstraete}}]{abreu2022}%
  \BibitemOpen
  \bibfield  {author} {\bibinfo {author} {\bibfnamefont {J.~C.}\ \bibnamefont
  {Abreu}}, \bibinfo {author} {\bibfnamefont {J.~P.}\ \bibnamefont {Nery}},
  \bibinfo {author} {\bibfnamefont {M.}~\bibnamefont {Giantomassi}}, \bibinfo
  {author} {\bibfnamefont {X.}~\bibnamefont {Gonze}},\ and\ \bibinfo {author}
  {\bibfnamefont {M.~J.}\ \bibnamefont {Verstraete}},\ }\bibfield  {title}
  {\bibinfo {title} {Spectroscopic signatures of nonpolarons~: the case of
  diamond},\ }\href@noop {} {\bibfield  {journal} {\bibinfo  {journal} {Phys.
  Chem. Chem. Phys.}\ }\textbf {\bibinfo {volume} {24}},\ \bibinfo {pages}
  {12580} (\bibinfo {year} {2022})}\BibitemShut {NoStop}%
\bibitem [{\citenamefont {Migdal}(1958)}]{Migdal}%
  \BibitemOpen
  \bibfield  {author} {\bibinfo {author} {\bibfnamefont {A.~B.}\ \bibnamefont
  {Migdal}},\ }\bibfield  {title} {\bibinfo {title} {Interaction between
  electrons and lattice vibrations in a normal metal},\ }\href@noop {}
  {\bibfield  {journal} {\bibinfo  {journal} {Zh. Eksp. Teor. Fiz.}\ }\textbf
  {\bibinfo {volume} {34}},\ \bibinfo {pages} {1438} (\bibinfo {year}
  {1958})}\BibitemShut {NoStop}%
\bibitem [{\citenamefont {Dyson}(1949)}]{Dyson}%
  \BibitemOpen
  \bibfield  {author} {\bibinfo {author} {\bibfnamefont {F.~J.}\ \bibnamefont
  {Dyson}},\ }\bibfield  {title} {\bibinfo {title} {The s matrix in quantum
  electrodynamics},\ }\href {https://doi.org/10.1103/PhysRev.75.1736}
  {\bibfield  {journal} {\bibinfo  {journal} {Phys. Rev.}\ }\textbf {\bibinfo
  {volume} {75}},\ \bibinfo {pages} {1736} (\bibinfo {year}
  {1949})}\BibitemShut {NoStop}%
\bibitem [{\citenamefont {Guzzo}\ \emph {et~al.}(2012)\citenamefont {Guzzo},
  \citenamefont {Kas}, \citenamefont {Sottile}, \citenamefont {Silly},
  \citenamefont {Sirotti}, \citenamefont {Rehr},\ and\ \citenamefont
  {Reining}}]{guzzo2012}%
  \BibitemOpen
  \bibfield  {author} {\bibinfo {author} {\bibfnamefont {M.}~\bibnamefont
  {Guzzo}}, \bibinfo {author} {\bibfnamefont {J.}~\bibnamefont {Kas}}, \bibinfo
  {author} {\bibfnamefont {F.}~\bibnamefont {Sottile}}, \bibinfo {author}
  {\bibfnamefont {M.}~\bibnamefont {Silly}}, \bibinfo {author} {\bibfnamefont
  {F.}~\bibnamefont {Sirotti}}, \bibinfo {author} {\bibfnamefont
  {J.}~\bibnamefont {Rehr}},\ and\ \bibinfo {author} {\bibfnamefont
  {L.}~\bibnamefont {Reining}},\ }\bibfield  {title} {\bibinfo {title}
  {Excitonic effects in x-ray absorption spectroscopy},\ }\href@noop {}
  {\bibfield  {journal} {\bibinfo  {journal} {European Physical Journal B}\
  }\textbf {\bibinfo {volume} {85}},\ \bibinfo {pages} {324} (\bibinfo {year}
  {2012})}\BibitemShut {NoStop}%
\bibitem [{\citenamefont {Caruso}\ and\ \citenamefont
  {Giustino}(2015)}]{caruso2015}%
  \BibitemOpen
  \bibfield  {author} {\bibinfo {author} {\bibfnamefont {F.}~\bibnamefont
  {Caruso}}\ and\ \bibinfo {author} {\bibfnamefont {F.}~\bibnamefont
  {Giustino}},\ }\bibfield  {title} {\bibinfo {title} {Ab initio calculation of
  the excitonic absorption spectrum of silicon},\ }\href@noop {} {\bibfield
  {journal} {\bibinfo  {journal} {Physical Review B: Condensed Matter and
  Materials Physics}\ }\textbf {\bibinfo {volume} {92}},\ \bibinfo {pages}
  {045123} (\bibinfo {year} {2015})}\BibitemShut {NoStop}%
\bibitem [{\citenamefont {Kas}\ \emph {et~al.}(2014)\citenamefont {Kas},
  \citenamefont {Rehr},\ and\ \citenamefont {Soininen}}]{kas2014cumulant}%
  \BibitemOpen
  \bibfield  {author} {\bibinfo {author} {\bibfnamefont {J.~J.}\ \bibnamefont
  {Kas}}, \bibinfo {author} {\bibfnamefont {J.~J.}\ \bibnamefont {Rehr}},\ and\
  \bibinfo {author} {\bibfnamefont {J.~A.}\ \bibnamefont {Soininen}},\
  }\bibfield  {title} {\bibinfo {title} {Cumulant expansion approach for the
  simulation of inelastic losses in x-ray absorption spectra},\ }\href@noop {}
  {\bibfield  {journal} {\bibinfo  {journal} {Physical Review B}\ }\textbf
  {\bibinfo {volume} {90}},\ \bibinfo {pages} {085112} (\bibinfo {year}
  {2014})}\BibitemShut {NoStop}%
\bibitem [{\citenamefont {Aryasetiawan}\ \emph {et~al.}(1996)\citenamefont
  {Aryasetiawan}, \citenamefont {Hedin},\ and\ \citenamefont
  {Karlsson}}]{aryasetiawan1996}%
  \BibitemOpen
  \bibfield  {author} {\bibinfo {author} {\bibfnamefont {F.}~\bibnamefont
  {Aryasetiawan}}, \bibinfo {author} {\bibfnamefont {L.}~\bibnamefont
  {Hedin}},\ and\ \bibinfo {author} {\bibfnamefont {K.}~\bibnamefont
  {Karlsson}},\ }\bibfield  {title} {\bibinfo {title} {Electronic structure of
  solids: The gw method},\ }\href@noop {} {\bibfield  {journal} {\bibinfo
  {journal} {Physical Review Letters}\ }\textbf {\bibinfo {volume} {77}},\
  \bibinfo {pages} {2268} (\bibinfo {year} {1996})}\BibitemShut {NoStop}%
\bibitem [{\citenamefont {Gumhalter}\ \emph {et~al.}(2016)\citenamefont
  {Gumhalter}, \citenamefont {Kov\'a\v{c}},\ and\ \citenamefont
  {Sundaram}}]{gumhalter2016cumulant}%
  \BibitemOpen
  \bibfield  {author} {\bibinfo {author} {\bibfnamefont {B.}~\bibnamefont
  {Gumhalter}}, \bibinfo {author} {\bibfnamefont {V.}~\bibnamefont
  {Kov\'a\v{c}}},\ and\ \bibinfo {author} {\bibfnamefont {B.}~\bibnamefont
  {Sundaram}},\ }\bibfield  {title} {\bibinfo {title} {Cumulant expansion
  methods for describing electron-phonon interactions in solids},\ }\href@noop
  {} {\bibfield  {journal} {\bibinfo  {journal} {Journal of Physics: Condensed
  Matter}\ }\textbf {\bibinfo {volume} {28}},\ \bibinfo {pages} {155601}
  (\bibinfo {year} {2016})}\BibitemShut {NoStop}%
\bibitem [{\citenamefont {Nery}\ \emph {et~al.}(2018)\citenamefont {Nery},
  \citenamefont {Allen}, \citenamefont {Antonius}, \citenamefont {Reining},
  \citenamefont {Miglio},\ and\ \citenamefont
  {Gonze}}]{nery2018quasiparticles}%
  \BibitemOpen
  \bibfield  {author} {\bibinfo {author} {\bibfnamefont {J.~P.}\ \bibnamefont
  {Nery}}, \bibinfo {author} {\bibfnamefont {P.~B.}\ \bibnamefont {Allen}},
  \bibinfo {author} {\bibfnamefont {G.}~\bibnamefont {Antonius}}, \bibinfo
  {author} {\bibfnamefont {L.}~\bibnamefont {Reining}}, \bibinfo {author}
  {\bibfnamefont {A.}~\bibnamefont {Miglio}},\ and\ \bibinfo {author}
  {\bibfnamefont {X.}~\bibnamefont {Gonze}},\ }\bibfield  {title} {\bibinfo
  {title} {Quasiparticles and phonon satellites in spectral functions of
  semiconductors and insulators: Cumulants applied to the full first-principles
  theory and the fröhlich polaron},\ }\href
  {https://doi.org/10.1103/PhysRevB.97.115145} {\bibfield  {journal} {\bibinfo
  {journal} {Physical Review B}\ }\textbf {\bibinfo {volume} {97}},\ \bibinfo
  {pages} {115145} (\bibinfo {year} {2018})}\BibitemShut {NoStop}%
\bibitem [{ver(2024)}]{verstraete2024polaronic}%
  \BibitemOpen
  \bibfield  {title} {\bibinfo {title} {Polaronic effects in diamond: Non-polar
  material studies},\ }\href@noop {} {\bibfield  {journal} {\bibinfo  {journal}
  {Physical Chemistry Chemical Physics}\ } (\bibinfo {year}
  {2024})}\BibitemShut {NoStop}%
\bibitem [{\citenamefont {Mishchenko}\ \emph {et~al.}(2000)\citenamefont
  {Mishchenko}, \citenamefont {Prokof’ev}, \citenamefont {Sakamoto},\ and\
  \citenamefont {Svistunov}}]{Mishchenko2000}%
  \BibitemOpen
  \bibfield  {author} {\bibinfo {author} {\bibfnamefont {A.}~\bibnamefont
  {Mishchenko}}, \bibinfo {author} {\bibfnamefont {N.}~\bibnamefont
  {Prokof’ev}}, \bibinfo {author} {\bibfnamefont {A.}~\bibnamefont
  {Sakamoto}},\ and\ \bibinfo {author} {\bibfnamefont {B.}~\bibnamefont
  {Svistunov}},\ }\bibfield  {title} {\bibinfo {title} {Electron-phonon
  interactions and coupled charge-lattice dynamics in a polaron model: A study
  with diagrammatic monte carlo},\ }\href
  {https://doi.org/10.1103/PhysRevB.62.6317} {\bibfield  {journal} {\bibinfo
  {journal} {Phys. Rev. B: Condens. Matter Mater. Phys.}\ }\textbf {\bibinfo
  {volume} {62}},\ \bibinfo {pages} {6317} (\bibinfo {year}
  {2000})}\BibitemShut {NoStop}%
\bibitem [{\citenamefont {Kubo}(1962)}]{Kubo1962}%
  \BibitemOpen
  \bibfield  {author} {\bibinfo {author} {\bibfnamefont {R.}~\bibnamefont
  {Kubo}},\ }\bibfield  {title} {\bibinfo {title} {Generalized cumulant
  expansion method},\ }\href {https://doi.org/10.1143/JPSJ.17.1100} {\bibfield
  {journal} {\bibinfo  {journal} {J. Phys. Soc. Jpn.}\ }\textbf {\bibinfo
  {volume} {17}},\ \bibinfo {pages} {1100} (\bibinfo {year}
  {1962})}\BibitemShut {NoStop}%
\bibitem [{\citenamefont {Gunnarsson}\ \emph {et~al.}(1994)\citenamefont
  {Gunnarsson}, \citenamefont {Meden},\ and\ \citenamefont
  {Schönhammer}}]{Gunnarsson1994}%
  \BibitemOpen
  \bibfield  {author} {\bibinfo {author} {\bibfnamefont {O.}~\bibnamefont
  {Gunnarsson}}, \bibinfo {author} {\bibfnamefont {V.}~\bibnamefont {Meden}},\
  and\ \bibinfo {author} {\bibfnamefont {K.}~\bibnamefont {Schönhammer}},\
  }\bibfield  {title} {\bibinfo {title} {Photoemission spectra of
  one-dimensional luttinger liquids},\ }\href
  {https://doi.org/10.1103/PhysRevB.50.10462} {\bibfield  {journal} {\bibinfo
  {journal} {Phys. Rev. B: Condens. Matter Mater. Phys.}\ }\textbf {\bibinfo
  {volume} {50}},\ \bibinfo {pages} {10462} (\bibinfo {year}
  {1994})}\BibitemShut {NoStop}%
\bibitem [{\citenamefont {Mahan}(2000)}]{Mahan2000}%
  \BibitemOpen
  \bibfield  {author} {\bibinfo {author} {\bibfnamefont {G.~D.}\ \bibnamefont
  {Mahan}},\ }\href@noop {} {\emph {\bibinfo {title} {Many-Particle Physics}}}\
  (\bibinfo  {publisher} {Springer},\ \bibinfo {address} {Boston},\ \bibinfo
  {year} {2000})\BibitemShut {NoStop}%
\bibitem [{\citenamefont {Antonius}\ \emph {et~al.}(2015)\citenamefont
  {Antonius}, \citenamefont {Ponc\'e}, \citenamefont {Lantagne-Hurtubise},
  \citenamefont {Auclair}, \citenamefont {Gonze},\ and\ \citenamefont
  {C\^ot\'e}}]{Antonius2015}%
  \BibitemOpen
  \bibfield  {author} {\bibinfo {author} {\bibfnamefont {G.}~\bibnamefont
  {Antonius}}, \bibinfo {author} {\bibfnamefont {S.}~\bibnamefont {Ponc\'e}},
  \bibinfo {author} {\bibfnamefont {E.}~\bibnamefont {Lantagne-Hurtubise}},
  \bibinfo {author} {\bibfnamefont {G.}~\bibnamefont {Auclair}}, \bibinfo
  {author} {\bibfnamefont {X.}~\bibnamefont {Gonze}},\ and\ \bibinfo {author}
  {\bibfnamefont {M.}~\bibnamefont {C\^ot\'e}},\ }\bibfield  {title} {\bibinfo
  {title} {Theory of the zero-point renormalization of the optical gap of
  diamond, silicon, and germanium},\ }\href
  {https://doi.org/10.1103/PhysRevB.92.085137} {\bibfield  {journal} {\bibinfo
  {journal} {Phys. Rev. B: Condens. Matter Mater. Phys.}\ }\textbf {\bibinfo
  {volume} {92}},\ \bibinfo {pages} {085137} (\bibinfo {year}
  {2015})}\BibitemShut {NoStop}%
\bibitem [{\citenamefont {Hedin}(1980)}]{Hedin1980}%
  \BibitemOpen
  \bibfield  {author} {\bibinfo {author} {\bibfnamefont {L.}~\bibnamefont
  {Hedin}},\ }\bibfield  {title} {\bibinfo {title} {Effects of recoil on
  shake-up spectra in metals},\ }\href
  {https://doi.org/10.1088/0031-8949/21/3-4/039} {\bibfield  {journal}
  {\bibinfo  {journal} {Physica Scripta}\ }\textbf {\bibinfo {volume} {21}},\
  \bibinfo {pages} {477–480} (\bibinfo {year} {1980})}\BibitemShut {NoStop}%
\bibitem [{\citenamefont {Guzzo}\ \emph {et~al.}(2011)\citenamefont {Guzzo},
  \citenamefont {Lani}, \citenamefont {Sottile}, \citenamefont {Romaniello},
  \citenamefont {Gatti}, \citenamefont {Kas}, \citenamefont {Rehr},
  \citenamefont {Silly}, \citenamefont {Sirotti},\ and\ \citenamefont
  {Reining}}]{Guzzo2011}%
  \BibitemOpen
  \bibfield  {author} {\bibinfo {author} {\bibfnamefont {M.}~\bibnamefont
  {Guzzo}}, \bibinfo {author} {\bibfnamefont {G.}~\bibnamefont {Lani}},
  \bibinfo {author} {\bibfnamefont {F.}~\bibnamefont {Sottile}}, \bibinfo
  {author} {\bibfnamefont {P.}~\bibnamefont {Romaniello}}, \bibinfo {author}
  {\bibfnamefont {M.}~\bibnamefont {Gatti}}, \bibinfo {author} {\bibfnamefont
  {J.~J.}\ \bibnamefont {Kas}}, \bibinfo {author} {\bibfnamefont {J.~J.}\
  \bibnamefont {Rehr}}, \bibinfo {author} {\bibfnamefont {M.~G.}\ \bibnamefont
  {Silly}}, \bibinfo {author} {\bibfnamefont {F.}~\bibnamefont {Sirotti}},\
  and\ \bibinfo {author} {\bibfnamefont {L.}~\bibnamefont {Reining}},\
  }\bibfield  {title} {\bibinfo {title} {Valence electron photoemission
  spectrum of semiconductors:ab initiodescription of multiple satellites},\
  }\bibfield  {journal} {\bibinfo  {journal} {Physical Review Letters}\
  }\textbf {\bibinfo {volume} {107}},\ \href
  {https://doi.org/10.1103/physrevlett.107.166401}
  {10.1103/physrevlett.107.166401} (\bibinfo {year} {2011})\BibitemShut
  {NoStop}%
\bibitem [{\citenamefont {Gumhalter}(2005)}]{Gumhalter2005}%
  \BibitemOpen
  \bibfield  {author} {\bibinfo {author} {\bibfnamefont {B.}~\bibnamefont
  {Gumhalter}},\ }\bibfield  {title} {\bibinfo {title} {Ultrafast dynamics and
  decoherence of quasiparticles in surface bands: Development of the
  formalism},\ }\bibfield  {journal} {\bibinfo  {journal} {Physical Review B}\
  }\textbf {\bibinfo {volume} {72}},\ \href
  {https://doi.org/10.1103/physrevb.72.165406} {10.1103/physrevb.72.165406}
  (\bibinfo {year} {2005})\BibitemShut {NoStop}%
\bibitem [{\citenamefont {Lihm}\ and\ \citenamefont {Poncé}(2025)}]{JMo2025}%
  \BibitemOpen
  \bibfield  {author} {\bibinfo {author} {\bibfnamefont {J.-M.}\ \bibnamefont
  {Lihm}}\ and\ \bibinfo {author} {\bibfnamefont {S.}~\bibnamefont {Poncé}},\
  }\href {https://doi.org/10.48550/ARXIV.2506.18139} {\bibinfo {title}
  {Beyond-quasiparticles transport with vertex correction: self-consistent
  ladder formalism for electron-phonon interactions}} (\bibinfo {year}
  {2025})\BibitemShut {NoStop}%
\bibitem [{\citenamefont {Zhou}\ and\ \citenamefont
  {Bernardi}(2019)}]{Zhou2019}%
  \BibitemOpen
  \bibfield  {author} {\bibinfo {author} {\bibfnamefont {J.-J.}\ \bibnamefont
  {Zhou}}\ and\ \bibinfo {author} {\bibfnamefont {M.}~\bibnamefont
  {Bernardi}},\ }\bibfield  {title} {\bibinfo {title} {Predicting charge
  transport in the presence of polarons: The beyond-quasiparticle regime in
  srtio3},\ }\bibfield  {journal} {\bibinfo  {journal} {Physical Review
  Research}\ }\textbf {\bibinfo {volume} {1}},\ \href
  {https://doi.org/10.1103/physrevresearch.1.033138}
  {10.1103/physrevresearch.1.033138} (\bibinfo {year} {2019})\BibitemShut
  {NoStop}%
\bibitem [{\citenamefont {Verdi}\ \emph {et~al.}(2017)\citenamefont {Verdi},
  \citenamefont {Caruso},\ and\ \citenamefont {Giustino}}]{Verdi2017}%
  \BibitemOpen
  \bibfield  {author} {\bibinfo {author} {\bibfnamefont {C.}~\bibnamefont
  {Verdi}}, \bibinfo {author} {\bibfnamefont {F.}~\bibnamefont {Caruso}},\ and\
  \bibinfo {author} {\bibfnamefont {F.}~\bibnamefont {Giustino}},\ }\bibfield
  {title} {\bibinfo {title} {Origin of the crossover from polarons to fermi
  liquids in transition metal oxides},\ }\bibfield  {journal} {\bibinfo
  {journal} {Nature Communications}\ }\textbf {\bibinfo {volume} {8}},\ \href
  {https://doi.org/10.1038/ncomms15769} {10.1038/ncomms15769} (\bibinfo {year}
  {2017})\BibitemShut {NoStop}%
\bibitem [{\citenamefont {Brooks}(1951)}]{Brooks1951}%
  \BibitemOpen
  \bibfield  {author} {\bibinfo {author} {\bibfnamefont {H.}~\bibnamefont
  {Brooks}},\ }\bibfield  {title} {\bibinfo {title} {Scattering by ionized
  impurities in semiconductors},\ }\href@noop {} {\bibfield  {journal}
  {\bibinfo  {journal} {Physical Review}\ }\textbf {\bibinfo {volume} {83}}
  (\bibinfo {year} {1951})}\BibitemShut {NoStop}%
\bibitem [{\citenamefont {Verstraete}\ \emph {et~al.}(2025)\citenamefont
  {Verstraete}, \citenamefont {Abreu}, \citenamefont {Allemand}, \citenamefont
  {Amadon}, \citenamefont {Antonius}, \citenamefont {Azizi}, \citenamefont
  {Baguet}, \citenamefont {Barat}, \citenamefont {Bastogne}, \citenamefont
  {Bejaud}, \citenamefont {Beuken}, \citenamefont {Bieder}, \citenamefont
  {Blanchet}, \citenamefont {Bottin}, \citenamefont {Bouchet}, \citenamefont
  {Bouquiaux}, \citenamefont {Bousquet}, \citenamefont {Boust}, \citenamefont
  {Brieuc}, \citenamefont {Brousseau-Couture}, \citenamefont {Bruneval},
  \citenamefont {Castellano}, \citenamefont {Castiel}, \citenamefont
  {Charraud}, \citenamefont {Clerouin}, \citenamefont {Cote}, \citenamefont
  {Duval}, \citenamefont {Gallo}, \citenamefont {Gendron}, \citenamefont
  {Geneste}, \citenamefont {Ghosez}, \citenamefont {Giantomassi}, \citenamefont
  {Gingras}, \citenamefont {Gomez-Ortiz}, \citenamefont {Gonze}, \citenamefont
  {Goudreault}, \citenamefont {Gruneis}, \citenamefont {Gupta}, \citenamefont
  {Guster}, \citenamefont {Hamann}, \citenamefont {He}, \citenamefont
  {Hellman}, \citenamefont {Holzwarth}, \citenamefont {Jollet}, \citenamefont
  {Kestener}, \citenamefont {Lygatsika}, \citenamefont {Nadeau}, \citenamefont
  {MacEnulty}, \citenamefont {Marazzi}, \citenamefont {Mignolet}, \citenamefont
  {O'Regan}, \citenamefont {Outerovitch}, \citenamefont {Paillard},
  \citenamefont {Petretto}, \citenamefont {Ponce}, \citenamefont {Ricci},
  \citenamefont {Rignanese}, \citenamefont {Rodriguez-Mayorga}, \citenamefont
  {Romero}, \citenamefont {Rostami}, \citenamefont {Royo}, \citenamefont
  {Sarraute}, \citenamefont {Sasani}, \citenamefont {Soubiran}, \citenamefont
  {Stengel}, \citenamefont {Tantardini}, \citenamefont {Torrent}, \citenamefont
  {Trinquet}, \citenamefont {Vasilchencko}, \citenamefont {Waroquiers},
  \citenamefont {Zabalo}, \citenamefont {Zadoks}, \citenamefont {Zhang},\ and\
  \citenamefont {Zwanziger}}]{Verstraete2025}%
  \BibitemOpen
  \bibfield  {author} {\bibinfo {author} {\bibfnamefont {M.~J.}\ \bibnamefont
  {Verstraete}}, \bibinfo {author} {\bibfnamefont {J.}~\bibnamefont {Abreu}},
  \bibinfo {author} {\bibfnamefont {G.~E.}\ \bibnamefont {Allemand}}, \bibinfo
  {author} {\bibfnamefont {B.}~\bibnamefont {Amadon}}, \bibinfo {author}
  {\bibfnamefont {G.}~\bibnamefont {Antonius}}, \bibinfo {author}
  {\bibfnamefont {M.}~\bibnamefont {Azizi}}, \bibinfo {author} {\bibfnamefont
  {L.}~\bibnamefont {Baguet}}, \bibinfo {author} {\bibfnamefont
  {C.}~\bibnamefont {Barat}}, \bibinfo {author} {\bibfnamefont
  {L.}~\bibnamefont {Bastogne}}, \bibinfo {author} {\bibfnamefont
  {R.}~\bibnamefont {Bejaud}}, \bibinfo {author} {\bibfnamefont {J.-M.}\
  \bibnamefont {Beuken}}, \bibinfo {author} {\bibfnamefont {J.}~\bibnamefont
  {Bieder}}, \bibinfo {author} {\bibfnamefont {A.}~\bibnamefont {Blanchet}},
  \bibinfo {author} {\bibfnamefont {F.}~\bibnamefont {Bottin}}, \bibinfo
  {author} {\bibfnamefont {J.}~\bibnamefont {Bouchet}}, \bibinfo {author}
  {\bibfnamefont {J.}~\bibnamefont {Bouquiaux}}, \bibinfo {author}
  {\bibfnamefont {E.}~\bibnamefont {Bousquet}}, \bibinfo {author}
  {\bibfnamefont {J.}~\bibnamefont {Boust}}, \bibinfo {author} {\bibfnamefont
  {F.}~\bibnamefont {Brieuc}}, \bibinfo {author} {\bibfnamefont
  {V.}~\bibnamefont {Brousseau-Couture}}, \bibinfo {author} {\bibfnamefont
  {N.~B.~F.}\ \bibnamefont {Bruneval}}, \bibinfo {author} {\bibfnamefont
  {A.}~\bibnamefont {Castellano}}, \bibinfo {author} {\bibfnamefont
  {E.}~\bibnamefont {Castiel}}, \bibinfo {author} {\bibfnamefont {J.-B.}\
  \bibnamefont {Charraud}}, \bibinfo {author} {\bibfnamefont {J.}~\bibnamefont
  {Clerouin}}, \bibinfo {author} {\bibfnamefont {M.}~\bibnamefont {Cote}},
  \bibinfo {author} {\bibfnamefont {C.}~\bibnamefont {Duval}}, \bibinfo
  {author} {\bibfnamefont {A.}~\bibnamefont {Gallo}}, \bibinfo {author}
  {\bibfnamefont {F.}~\bibnamefont {Gendron}}, \bibinfo {author} {\bibfnamefont
  {G.}~\bibnamefont {Geneste}}, \bibinfo {author} {\bibfnamefont
  {P.}~\bibnamefont {Ghosez}}, \bibinfo {author} {\bibfnamefont
  {M.}~\bibnamefont {Giantomassi}}, \bibinfo {author} {\bibfnamefont
  {O.}~\bibnamefont {Gingras}}, \bibinfo {author} {\bibfnamefont
  {F.}~\bibnamefont {Gomez-Ortiz}}, \bibinfo {author} {\bibfnamefont
  {X.}~\bibnamefont {Gonze}}, \bibinfo {author} {\bibfnamefont {F.~A.}\
  \bibnamefont {Goudreault}}, \bibinfo {author} {\bibfnamefont
  {A.}~\bibnamefont {Gruneis}}, \bibinfo {author} {\bibfnamefont
  {R.}~\bibnamefont {Gupta}}, \bibinfo {author} {\bibfnamefont
  {B.}~\bibnamefont {Guster}}, \bibinfo {author} {\bibfnamefont {D.~R.}\
  \bibnamefont {Hamann}}, \bibinfo {author} {\bibfnamefont {X.}~\bibnamefont
  {He}}, \bibinfo {author} {\bibfnamefont {O.}~\bibnamefont {Hellman}},
  \bibinfo {author} {\bibfnamefont {N.}~\bibnamefont {Holzwarth}}, \bibinfo
  {author} {\bibfnamefont {F.}~\bibnamefont {Jollet}}, \bibinfo {author}
  {\bibfnamefont {P.}~\bibnamefont {Kestener}}, \bibinfo {author}
  {\bibfnamefont {I.-M.}\ \bibnamefont {Lygatsika}}, \bibinfo {author}
  {\bibfnamefont {O.}~\bibnamefont {Nadeau}}, \bibinfo {author} {\bibfnamefont
  {L.}~\bibnamefont {MacEnulty}}, \bibinfo {author} {\bibfnamefont
  {E.}~\bibnamefont {Marazzi}}, \bibinfo {author} {\bibfnamefont
  {M.}~\bibnamefont {Mignolet}}, \bibinfo {author} {\bibfnamefont {D.~D.}\
  \bibnamefont {O'Regan}}, \bibinfo {author} {\bibfnamefont {R.}~\bibnamefont
  {Outerovitch}}, \bibinfo {author} {\bibfnamefont {C.}~\bibnamefont
  {Paillard}}, \bibinfo {author} {\bibfnamefont {G.}~\bibnamefont {Petretto}},
  \bibinfo {author} {\bibfnamefont {S.}~\bibnamefont {Ponce}}, \bibinfo
  {author} {\bibfnamefont {F.}~\bibnamefont {Ricci}}, \bibinfo {author}
  {\bibfnamefont {G.-M.}\ \bibnamefont {Rignanese}}, \bibinfo {author}
  {\bibfnamefont {M.}~\bibnamefont {Rodriguez-Mayorga}}, \bibinfo {author}
  {\bibfnamefont {A.~H.}\ \bibnamefont {Romero}}, \bibinfo {author}
  {\bibfnamefont {S.}~\bibnamefont {Rostami}}, \bibinfo {author} {\bibfnamefont
  {M.}~\bibnamefont {Royo}}, \bibinfo {author} {\bibfnamefont {M.}~\bibnamefont
  {Sarraute}}, \bibinfo {author} {\bibfnamefont {A.}~\bibnamefont {Sasani}},
  \bibinfo {author} {\bibfnamefont {F.}~\bibnamefont {Soubiran}}, \bibinfo
  {author} {\bibfnamefont {M.}~\bibnamefont {Stengel}}, \bibinfo {author}
  {\bibfnamefont {C.}~\bibnamefont {Tantardini}}, \bibinfo {author}
  {\bibfnamefont {M.}~\bibnamefont {Torrent}}, \bibinfo {author} {\bibfnamefont
  {V.}~\bibnamefont {Trinquet}}, \bibinfo {author} {\bibfnamefont
  {V.}~\bibnamefont {Vasilchencko}}, \bibinfo {author} {\bibfnamefont
  {D.}~\bibnamefont {Waroquiers}}, \bibinfo {author} {\bibfnamefont
  {A.}~\bibnamefont {Zabalo}}, \bibinfo {author} {\bibfnamefont
  {A.}~\bibnamefont {Zadoks}}, \bibinfo {author} {\bibfnamefont
  {H.}~\bibnamefont {Zhang}},\ and\ \bibinfo {author} {\bibfnamefont
  {J.}~\bibnamefont {Zwanziger}},\ }\href
  {https://doi.org/10.48550/ARXIV.2507.08578} {\bibinfo {title} {Abinit 2025:
  New capabilities for the predictive modeling of solids and nanomaterials}}
  (\bibinfo {year} {2025})\BibitemShut {NoStop}%
\bibitem [{\citenamefont {Gonze}\ \emph {et~al.}(2020)\citenamefont {Gonze},
  \citenamefont {Amadon}, \citenamefont {Antonius}, \citenamefont {Arnardi},
  \citenamefont {Baguet}, \citenamefont {Beuken}, \citenamefont {Bieder},
  \citenamefont {Bottin}, \citenamefont {Bouchet}, \citenamefont {Bousquet},
  \citenamefont {Brouwer}, \citenamefont {Bruneval}, \citenamefont {Brunin},
  \citenamefont {Cavignac}, \citenamefont {Charraud}, \citenamefont {Chen},
  \citenamefont {C\^oté}, \citenamefont {Cottenier}, \citenamefont {Denier},
  \citenamefont {Geneste}, \citenamefont {Ghosez}, \citenamefont {Giantomassi},
  \citenamefont {Gillet}, \citenamefont {Gingras}, \citenamefont {Hamann},
  \citenamefont {Hautier}, \citenamefont {He}, \citenamefont {Helbig},
  \citenamefont {Holzwarth}, \citenamefont {Jia}, \citenamefont {Jollet},
  \citenamefont {Lafargue-Dit-Hauret}, \citenamefont {Lejaeghere},
  \citenamefont {Marques}, \citenamefont {Martin}, \citenamefont {Martins},
  \citenamefont {Miranda}, \citenamefont {Naccarato}, \citenamefont {Persson},
  \citenamefont {Petretto}, \citenamefont {Planes}, \citenamefont {Pouillon},
  \citenamefont {Prokhorenko}, \citenamefont {Ricci}, \citenamefont
  {Rignanese}, \citenamefont {Romero}, \citenamefont {Schmitt}, \citenamefont
  {Torrent}, \citenamefont {van Setten}, \citenamefont {Van~Troeye},
  \citenamefont {Verstraete}, \citenamefont {Zérah},\ and\ \citenamefont
  {Zwanziger}}]{Gonze2020}%
  \BibitemOpen
  \bibfield  {author} {\bibinfo {author} {\bibfnamefont {X.}~\bibnamefont
  {Gonze}}, \bibinfo {author} {\bibfnamefont {B.}~\bibnamefont {Amadon}},
  \bibinfo {author} {\bibfnamefont {G.}~\bibnamefont {Antonius}}, \bibinfo
  {author} {\bibfnamefont {F.}~\bibnamefont {Arnardi}}, \bibinfo {author}
  {\bibfnamefont {L.}~\bibnamefont {Baguet}}, \bibinfo {author} {\bibfnamefont
  {J.-M.}\ \bibnamefont {Beuken}}, \bibinfo {author} {\bibfnamefont
  {J.}~\bibnamefont {Bieder}}, \bibinfo {author} {\bibfnamefont
  {F.}~\bibnamefont {Bottin}}, \bibinfo {author} {\bibfnamefont
  {J.}~\bibnamefont {Bouchet}}, \bibinfo {author} {\bibfnamefont
  {E.}~\bibnamefont {Bousquet}}, \bibinfo {author} {\bibfnamefont
  {N.}~\bibnamefont {Brouwer}}, \bibinfo {author} {\bibfnamefont
  {F.}~\bibnamefont {Bruneval}}, \bibinfo {author} {\bibfnamefont
  {G.}~\bibnamefont {Brunin}}, \bibinfo {author} {\bibfnamefont
  {T.}~\bibnamefont {Cavignac}}, \bibinfo {author} {\bibfnamefont {J.-B.}\
  \bibnamefont {Charraud}}, \bibinfo {author} {\bibfnamefont {W.}~\bibnamefont
  {Chen}}, \bibinfo {author} {\bibfnamefont {M.}~\bibnamefont {C\^oté}},
  \bibinfo {author} {\bibfnamefont {S.}~\bibnamefont {Cottenier}}, \bibinfo
  {author} {\bibfnamefont {J.}~\bibnamefont {Denier}}, \bibinfo {author}
  {\bibfnamefont {G.}~\bibnamefont {Geneste}}, \bibinfo {author} {\bibfnamefont
  {P.}~\bibnamefont {Ghosez}}, \bibinfo {author} {\bibfnamefont
  {M.}~\bibnamefont {Giantomassi}}, \bibinfo {author} {\bibfnamefont
  {Y.}~\bibnamefont {Gillet}}, \bibinfo {author} {\bibfnamefont
  {O.}~\bibnamefont {Gingras}}, \bibinfo {author} {\bibfnamefont {D.~R.}\
  \bibnamefont {Hamann}}, \bibinfo {author} {\bibfnamefont {G.}~\bibnamefont
  {Hautier}}, \bibinfo {author} {\bibfnamefont {X.}~\bibnamefont {He}},
  \bibinfo {author} {\bibfnamefont {N.}~\bibnamefont {Helbig}}, \bibinfo
  {author} {\bibfnamefont {N.}~\bibnamefont {Holzwarth}}, \bibinfo {author}
  {\bibfnamefont {Y.}~\bibnamefont {Jia}}, \bibinfo {author} {\bibfnamefont
  {F.}~\bibnamefont {Jollet}}, \bibinfo {author} {\bibfnamefont
  {W.}~\bibnamefont {Lafargue-Dit-Hauret}}, \bibinfo {author} {\bibfnamefont
  {K.}~\bibnamefont {Lejaeghere}}, \bibinfo {author} {\bibfnamefont {M.~A.}\
  \bibnamefont {Marques}}, \bibinfo {author} {\bibfnamefont {A.}~\bibnamefont
  {Martin}}, \bibinfo {author} {\bibfnamefont {C.}~\bibnamefont {Martins}},
  \bibinfo {author} {\bibfnamefont {H.~P.}\ \bibnamefont {Miranda}}, \bibinfo
  {author} {\bibfnamefont {F.}~\bibnamefont {Naccarato}}, \bibinfo {author}
  {\bibfnamefont {K.}~\bibnamefont {Persson}}, \bibinfo {author} {\bibfnamefont
  {G.}~\bibnamefont {Petretto}}, \bibinfo {author} {\bibfnamefont
  {V.}~\bibnamefont {Planes}}, \bibinfo {author} {\bibfnamefont
  {Y.}~\bibnamefont {Pouillon}}, \bibinfo {author} {\bibfnamefont
  {S.}~\bibnamefont {Prokhorenko}}, \bibinfo {author} {\bibfnamefont
  {F.}~\bibnamefont {Ricci}}, \bibinfo {author} {\bibfnamefont {G.-M.}\
  \bibnamefont {Rignanese}}, \bibinfo {author} {\bibfnamefont {A.~H.}\
  \bibnamefont {Romero}}, \bibinfo {author} {\bibfnamefont {M.~M.}\
  \bibnamefont {Schmitt}}, \bibinfo {author} {\bibfnamefont {M.}~\bibnamefont
  {Torrent}}, \bibinfo {author} {\bibfnamefont {M.~J.}\ \bibnamefont {van
  Setten}}, \bibinfo {author} {\bibfnamefont {B.}~\bibnamefont {Van~Troeye}},
  \bibinfo {author} {\bibfnamefont {M.~J.}\ \bibnamefont {Verstraete}},
  \bibinfo {author} {\bibfnamefont {G.}~\bibnamefont {Zérah}},\ and\ \bibinfo
  {author} {\bibfnamefont {J.~W.}\ \bibnamefont {Zwanziger}},\ }\bibfield
  {title} {\bibinfo {title} {The abinitproject: Impact, environment and recent
  developments},\ }\href {https://doi.org/10.1016/j.cpc.2019.107042} {\bibfield
   {journal} {\bibinfo  {journal} {Computer Physics Communications}\ }\textbf
  {\bibinfo {volume} {248}},\ \bibinfo {pages} {107042} (\bibinfo {year}
  {2020})}\BibitemShut {NoStop}%
\bibitem [{\citenamefont {Romero}\ \emph {et~al.}(2020)\citenamefont {Romero},
  \citenamefont {Allan}, \citenamefont {Amadon}, \citenamefont {Antonius},
  \citenamefont {Applencourt}, \citenamefont {Baguet}, \citenamefont {Bieder},
  \citenamefont {Bottin}, \citenamefont {Bouchet}, \citenamefont {Bousquet},
  \citenamefont {Bruneval}, \citenamefont {Brunin}, \citenamefont {Caliste},
  \citenamefont {C\^oté}, \citenamefont {Denier}, \citenamefont {Dreyer},
  \citenamefont {Ghosez}, \citenamefont {Giantomassi}, \citenamefont {Gillet},
  \citenamefont {Gingras}, \citenamefont {Hamann}, \citenamefont {Hautier},
  \citenamefont {Jollet}, \citenamefont {Jomard}, \citenamefont {Martin},
  \citenamefont {Miranda}, \citenamefont {Naccarato}, \citenamefont {Petretto},
  \citenamefont {Pike}, \citenamefont {Planes}, \citenamefont {Prokhorenko},
  \citenamefont {Rangel}, \citenamefont {Ricci}, \citenamefont {Rignanese},
  \citenamefont {Royo}, \citenamefont {Stengel}, \citenamefont {Torrent},
  \citenamefont {van Setten}, \citenamefont {Van~Troeye}, \citenamefont
  {Verstraete}, \citenamefont {Wiktor}, \citenamefont {Zwanziger},\ and\
  \citenamefont {Gonze}}]{Romero2020}%
  \BibitemOpen
  \bibfield  {author} {\bibinfo {author} {\bibfnamefont {A.~H.}\ \bibnamefont
  {Romero}}, \bibinfo {author} {\bibfnamefont {D.~C.}\ \bibnamefont {Allan}},
  \bibinfo {author} {\bibfnamefont {B.}~\bibnamefont {Amadon}}, \bibinfo
  {author} {\bibfnamefont {G.}~\bibnamefont {Antonius}}, \bibinfo {author}
  {\bibfnamefont {T.}~\bibnamefont {Applencourt}}, \bibinfo {author}
  {\bibfnamefont {L.}~\bibnamefont {Baguet}}, \bibinfo {author} {\bibfnamefont
  {J.}~\bibnamefont {Bieder}}, \bibinfo {author} {\bibfnamefont
  {F.}~\bibnamefont {Bottin}}, \bibinfo {author} {\bibfnamefont
  {J.}~\bibnamefont {Bouchet}}, \bibinfo {author} {\bibfnamefont
  {E.}~\bibnamefont {Bousquet}}, \bibinfo {author} {\bibfnamefont
  {F.}~\bibnamefont {Bruneval}}, \bibinfo {author} {\bibfnamefont
  {G.}~\bibnamefont {Brunin}}, \bibinfo {author} {\bibfnamefont
  {D.}~\bibnamefont {Caliste}}, \bibinfo {author} {\bibfnamefont
  {M.}~\bibnamefont {C\^oté}}, \bibinfo {author} {\bibfnamefont
  {J.}~\bibnamefont {Denier}}, \bibinfo {author} {\bibfnamefont
  {C.}~\bibnamefont {Dreyer}}, \bibinfo {author} {\bibfnamefont
  {P.}~\bibnamefont {Ghosez}}, \bibinfo {author} {\bibfnamefont
  {M.}~\bibnamefont {Giantomassi}}, \bibinfo {author} {\bibfnamefont
  {Y.}~\bibnamefont {Gillet}}, \bibinfo {author} {\bibfnamefont
  {O.}~\bibnamefont {Gingras}}, \bibinfo {author} {\bibfnamefont {D.~R.}\
  \bibnamefont {Hamann}}, \bibinfo {author} {\bibfnamefont {G.}~\bibnamefont
  {Hautier}}, \bibinfo {author} {\bibfnamefont {F.}~\bibnamefont {Jollet}},
  \bibinfo {author} {\bibfnamefont {G.}~\bibnamefont {Jomard}}, \bibinfo
  {author} {\bibfnamefont {A.}~\bibnamefont {Martin}}, \bibinfo {author}
  {\bibfnamefont {H.~P.~C.}\ \bibnamefont {Miranda}}, \bibinfo {author}
  {\bibfnamefont {F.}~\bibnamefont {Naccarato}}, \bibinfo {author}
  {\bibfnamefont {G.}~\bibnamefont {Petretto}}, \bibinfo {author}
  {\bibfnamefont {N.~A.}\ \bibnamefont {Pike}}, \bibinfo {author}
  {\bibfnamefont {V.}~\bibnamefont {Planes}}, \bibinfo {author} {\bibfnamefont
  {S.}~\bibnamefont {Prokhorenko}}, \bibinfo {author} {\bibfnamefont
  {T.}~\bibnamefont {Rangel}}, \bibinfo {author} {\bibfnamefont
  {F.}~\bibnamefont {Ricci}}, \bibinfo {author} {\bibfnamefont {G.-M.}\
  \bibnamefont {Rignanese}}, \bibinfo {author} {\bibfnamefont {M.}~\bibnamefont
  {Royo}}, \bibinfo {author} {\bibfnamefont {M.}~\bibnamefont {Stengel}},
  \bibinfo {author} {\bibfnamefont {M.}~\bibnamefont {Torrent}}, \bibinfo
  {author} {\bibfnamefont {M.~J.}\ \bibnamefont {van Setten}}, \bibinfo
  {author} {\bibfnamefont {B.}~\bibnamefont {Van~Troeye}}, \bibinfo {author}
  {\bibfnamefont {M.~J.}\ \bibnamefont {Verstraete}}, \bibinfo {author}
  {\bibfnamefont {J.}~\bibnamefont {Wiktor}}, \bibinfo {author} {\bibfnamefont
  {J.~W.}\ \bibnamefont {Zwanziger}},\ and\ \bibinfo {author} {\bibfnamefont
  {X.}~\bibnamefont {Gonze}},\ }\bibfield  {title} {\bibinfo {title} {Abinit:
  Overview and focus on selected capabilities},\ }\bibfield  {journal}
  {\bibinfo  {journal} {The Journal of Chemical Physics}\ }\textbf {\bibinfo
  {volume} {152}},\ \href {https://doi.org/10.1063/1.5144261}
  {10.1063/1.5144261} (\bibinfo {year} {2020})\BibitemShut {NoStop}%
\bibitem [{\citenamefont {Brunin}\ \emph
  {et~al.}(2020{\natexlab{a}})\citenamefont {Brunin}, \citenamefont {Miranda},
  \citenamefont {Giantomassi}, \citenamefont {Royo}, \citenamefont {Stengel},
  \citenamefont {Verstraete}, \citenamefont {Gonze}, \citenamefont
  {Rignanese},\ and\ \citenamefont {Hautier}}]{Brunin2020}%
  \BibitemOpen
  \bibfield  {author} {\bibinfo {author} {\bibfnamefont {G.}~\bibnamefont
  {Brunin}}, \bibinfo {author} {\bibfnamefont {H.~P.~C.}\ \bibnamefont
  {Miranda}}, \bibinfo {author} {\bibfnamefont {M.}~\bibnamefont
  {Giantomassi}}, \bibinfo {author} {\bibfnamefont {M.}~\bibnamefont {Royo}},
  \bibinfo {author} {\bibfnamefont {M.}~\bibnamefont {Stengel}}, \bibinfo
  {author} {\bibfnamefont {M.~J.}\ \bibnamefont {Verstraete}}, \bibinfo
  {author} {\bibfnamefont {X.}~\bibnamefont {Gonze}}, \bibinfo {author}
  {\bibfnamefont {G.-M.}\ \bibnamefont {Rignanese}},\ and\ \bibinfo {author}
  {\bibfnamefont {G.}~\bibnamefont {Hautier}},\ }\bibfield  {title} {\bibinfo
  {title} {Electron-phonon beyond fr\"{o}hlich: Dynamical quadrupoles in polar
  and covalent solids},\ }\bibfield  {journal} {\bibinfo  {journal} {Physical
  Review Letters}\ }\textbf {\bibinfo {volume} {125}},\ \href
  {https://doi.org/10.1103/physrevlett.125.136601}
  {10.1103/physrevlett.125.136601} (\bibinfo {year}
  {2020}{\natexlab{a}})\BibitemShut {NoStop}%
\bibitem [{\citenamefont {Brunin}\ \emph
  {et~al.}(2020{\natexlab{b}})\citenamefont {Brunin}, \citenamefont {Miranda},
  \citenamefont {Giantomassi}, \citenamefont {Royo}, \citenamefont {Stengel},
  \citenamefont {Verstraete}, \citenamefont {Gonze}, \citenamefont
  {Rignanese},\ and\ \citenamefont {Hautier}}]{Brunin2020prb}%
  \BibitemOpen
  \bibfield  {author} {\bibinfo {author} {\bibfnamefont {G.}~\bibnamefont
  {Brunin}}, \bibinfo {author} {\bibfnamefont {H.~P.~C.}\ \bibnamefont
  {Miranda}}, \bibinfo {author} {\bibfnamefont {M.}~\bibnamefont
  {Giantomassi}}, \bibinfo {author} {\bibfnamefont {M.}~\bibnamefont {Royo}},
  \bibinfo {author} {\bibfnamefont {M.}~\bibnamefont {Stengel}}, \bibinfo
  {author} {\bibfnamefont {M.~J.}\ \bibnamefont {Verstraete}}, \bibinfo
  {author} {\bibfnamefont {X.}~\bibnamefont {Gonze}}, \bibinfo {author}
  {\bibfnamefont {G.-M.}\ \bibnamefont {Rignanese}},\ and\ \bibinfo {author}
  {\bibfnamefont {G.}~\bibnamefont {Hautier}},\ }\bibfield  {title} {\bibinfo
  {title} {Phonon-limited electron mobility in si, gaas, and gap with exact
  treatment of dynamical quadrupoles},\ }\bibfield  {journal} {\bibinfo
  {journal} {Physical Review B}\ }\textbf {\bibinfo {volume} {102}},\ \href
  {https://doi.org/10.1103/physrevb.102.094308} {10.1103/physrevb.102.094308}
  (\bibinfo {year} {2020}{\natexlab{b}})\BibitemShut {NoStop}%
\bibitem [{\citenamefont {Eiguren}\ and\ \citenamefont
  {Ambrosch-Draxl}(2008)}]{Eiguren2008}%
  \BibitemOpen
  \bibfield  {author} {\bibinfo {author} {\bibfnamefont {A.}~\bibnamefont
  {Eiguren}}\ and\ \bibinfo {author} {\bibfnamefont {C.}~\bibnamefont
  {Ambrosch-Draxl}},\ }\bibfield  {title} {\bibinfo {title} {Wannier
  interpolation scheme for phonon-induced potentials: Application to bulk
  mgb$_2$, w, and the (1×1) h-covered w(110) surface},\ }\bibfield  {journal}
  {\bibinfo  {journal} {Physical Review B}\ }\textbf {\bibinfo {volume} {78}},\
  \href {https://doi.org/10.1103/physrevb.78.045124}
  {10.1103/physrevb.78.045124} (\bibinfo {year} {2008})\BibitemShut {NoStop}%
\bibitem [{\citenamefont {Levinshtein}\ \emph {et~al.}(2001)\citenamefont
  {Levinshtein}, \citenamefont {Rumyantsev},\ and\ \citenamefont
  {Shur}}]{Levinshtein2001-va}%
  \BibitemOpen
  \bibinfo {editor} {\bibfnamefont {M.~E.}\ \bibnamefont {Levinshtein}},
  \bibinfo {editor} {\bibfnamefont {S.~L.}\ \bibnamefont {Rumyantsev}},\ and\
  \bibinfo {editor} {\bibfnamefont {M.~S.}\ \bibnamefont {Shur}},\ eds.,\
  \href@noop {} {\emph {\bibinfo {title} {Properties of Advanced Semiconductor
  Materials}}}\ (\bibinfo  {publisher} {John Wiley \& Sons},\ \bibinfo
  {address} {Nashville, TN},\ \bibinfo {year} {2001})\BibitemShut {NoStop}%
\bibitem [{\citenamefont {Sio}\ \emph {et~al.}(2019)\citenamefont {Sio},
  \citenamefont {Verdi}, \citenamefont {Poncé},\ and\ \citenamefont
  {Giustino}}]{Sio2019}%
  \BibitemOpen
  \bibfield  {author} {\bibinfo {author} {\bibfnamefont {W.~H.}\ \bibnamefont
  {Sio}}, \bibinfo {author} {\bibfnamefont {C.}~\bibnamefont {Verdi}}, \bibinfo
  {author} {\bibfnamefont {S.}~\bibnamefont {Poncé}},\ and\ \bibinfo {author}
  {\bibfnamefont {F.}~\bibnamefont {Giustino}},\ }\bibfield  {title} {\bibinfo
  {title} {Ab initiotheory of polarons: Formalism and applications},\
  }\bibfield  {journal} {\bibinfo  {journal} {Physical Review B}\ }\textbf
  {\bibinfo {volume} {99}},\ \href {https://doi.org/10.1103/physrevb.99.235139}
  {10.1103/physrevb.99.235139} (\bibinfo {year} {2019})\BibitemShut {NoStop}%
\bibitem [{\citenamefont {Peeters}\ and\ \citenamefont
  {Devreese}(1985)}]{Peeters1985}%
  \BibitemOpen
  \bibfield  {author} {\bibinfo {author} {\bibfnamefont {F.~M.}\ \bibnamefont
  {Peeters}}\ and\ \bibinfo {author} {\bibfnamefont {J.~T.}\ \bibnamefont
  {Devreese}},\ }\bibfield  {title} {\bibinfo {title} {Acoustical polaron in
  three dimensions: The ground-state energy and the self-trapping transition},\
  }\href {https://doi.org/10.1103/physrevb.32.3515} {\bibfield  {journal}
  {\bibinfo  {journal} {Physical Review B}\ }\textbf {\bibinfo {volume} {32}},\
  \bibinfo {pages} {3515–3521} (\bibinfo {year} {1985})}\BibitemShut
  {NoStop}%
\bibitem [{\citenamefont {Jacoboni}\ \emph {et~al.}(1977)\citenamefont
  {Jacoboni}, \citenamefont {Canali}, \citenamefont {Ottaviani},\ and\
  \citenamefont {Alberigi~Quaranta}}]{Jacoboni1977}%
  \BibitemOpen
  \bibfield  {author} {\bibinfo {author} {\bibfnamefont {C.}~\bibnamefont
  {Jacoboni}}, \bibinfo {author} {\bibfnamefont {C.}~\bibnamefont {Canali}},
  \bibinfo {author} {\bibfnamefont {G.}~\bibnamefont {Ottaviani}},\ and\
  \bibinfo {author} {\bibfnamefont {A.}~\bibnamefont {Alberigi~Quaranta}},\
  }\bibfield  {title} {\bibinfo {title} {A review of some charge transport
  properties of silicon},\ }\href
  {https://doi.org/10.1016/0038-1101(77)90054-5} {\bibfield  {journal}
  {\bibinfo  {journal} {Solid-State Electronics}\ }\textbf {\bibinfo {volume}
  {20}},\ \bibinfo {pages} {77–89} (\bibinfo {year} {1977})}\BibitemShut
  {NoStop}%
\bibitem [{\citenamefont {Mousty}\ \emph {et~al.}(1974)\citenamefont {Mousty},
  \citenamefont {Ostoja},\ and\ \citenamefont {Passari}}]{Mousty1974}%
  \BibitemOpen
  \bibfield  {author} {\bibinfo {author} {\bibfnamefont {F.}~\bibnamefont
  {Mousty}}, \bibinfo {author} {\bibfnamefont {P.}~\bibnamefont {Ostoja}},\
  and\ \bibinfo {author} {\bibfnamefont {L.}~\bibnamefont {Passari}},\
  }\bibfield  {title} {\bibinfo {title} {Relationship between resistivity and
  phosphorus concentration in silicon},\ }\href
  {https://doi.org/10.1063/1.1663091} {\bibfield  {journal} {\bibinfo
  {journal} {Journal of Applied Physics}\ }\textbf {\bibinfo {volume} {45}},\
  \bibinfo {pages} {4576–4580} (\bibinfo {year} {1974})}\BibitemShut
  {NoStop}%
\bibitem [{\citenamefont {Norton}\ \emph {et~al.}(1973)\citenamefont {Norton},
  \citenamefont {Braggins},\ and\ \citenamefont {Levinstein}}]{Norton1973}%
  \BibitemOpen
  \bibfield  {author} {\bibinfo {author} {\bibfnamefont {P.}~\bibnamefont
  {Norton}}, \bibinfo {author} {\bibfnamefont {T.}~\bibnamefont {Braggins}},\
  and\ \bibinfo {author} {\bibfnamefont {H.}~\bibnamefont {Levinstein}},\
  }\bibfield  {title} {\bibinfo {title} {Impurity and lattice scattering
  parameters as determined from hall and mobility analysis in silicon},\ }\href
  {https://doi.org/10.1103/physrevb.8.5632} {\bibfield  {journal} {\bibinfo
  {journal} {Physical Review B}\ }\textbf {\bibinfo {volume} {8}},\ \bibinfo
  {pages} {5632–5653} (\bibinfo {year} {1973})}\BibitemShut {NoStop}%
\bibitem [{\citenamefont {Ottaviani}\ \emph {et~al.}(1975)\citenamefont
  {Ottaviani}, \citenamefont {Reggiani}, \citenamefont {Canali}, \citenamefont
  {Nava},\ and\ \citenamefont {Alberigi-Quaranta}}]{Ottaviani1975}%
  \BibitemOpen
  \bibfield  {author} {\bibinfo {author} {\bibfnamefont {G.}~\bibnamefont
  {Ottaviani}}, \bibinfo {author} {\bibfnamefont {L.}~\bibnamefont {Reggiani}},
  \bibinfo {author} {\bibfnamefont {C.}~\bibnamefont {Canali}}, \bibinfo
  {author} {\bibfnamefont {F.}~\bibnamefont {Nava}},\ and\ \bibinfo {author}
  {\bibfnamefont {A.}~\bibnamefont {Alberigi-Quaranta}},\ }\bibfield  {title}
  {\bibinfo {title} {Hole drift velocity in silicon},\ }\href
  {https://doi.org/10.1103/physrevb.12.3318} {\bibfield  {journal} {\bibinfo
  {journal} {Physical Review B}\ }\textbf {\bibinfo {volume} {12}},\ \bibinfo
  {pages} {3318–3329} (\bibinfo {year} {1975})}\BibitemShut {NoStop}%
\bibitem [{\citenamefont {Logan}\ and\ \citenamefont
  {Peters}(1960)}]{Logan1960}%
  \BibitemOpen
  \bibfield  {author} {\bibinfo {author} {\bibfnamefont {R.~A.}\ \bibnamefont
  {Logan}}\ and\ \bibinfo {author} {\bibfnamefont {A.~J.}\ \bibnamefont
  {Peters}},\ }\bibfield  {title} {\bibinfo {title} {Impurity effects upon
  mobility in silicon},\ }\href {https://doi.org/10.1063/1.1735385} {\bibfield
  {journal} {\bibinfo  {journal} {Journal of Applied Physics}\ }\textbf
  {\bibinfo {volume} {31}},\ \bibinfo {pages} {122–124} (\bibinfo {year}
  {1960})}\BibitemShut {NoStop}%
\bibitem [{\citenamefont {Jacoboni}\ \emph {et~al.}(1981)\citenamefont
  {Jacoboni}, \citenamefont {Nava}, \citenamefont {Canali},\ and\ \citenamefont
  {Ottaviani}}]{Jacoboni1981}%
  \BibitemOpen
  \bibfield  {author} {\bibinfo {author} {\bibfnamefont {C.}~\bibnamefont
  {Jacoboni}}, \bibinfo {author} {\bibfnamefont {F.}~\bibnamefont {Nava}},
  \bibinfo {author} {\bibfnamefont {C.}~\bibnamefont {Canali}},\ and\ \bibinfo
  {author} {\bibfnamefont {G.}~\bibnamefont {Ottaviani}},\ }\bibfield  {title}
  {\bibinfo {title} {Electron drift velocity and diffusivity in germanium},\
  }\href {https://doi.org/10.1103/physrevb.24.1014} {\bibfield  {journal}
  {\bibinfo  {journal} {Physical Review B}\ }\textbf {\bibinfo {volume} {24}},\
  \bibinfo {pages} {1014–1026} (\bibinfo {year} {1981})}\BibitemShut
  {NoStop}%
\bibitem [{\citenamefont {Debye}\ and\ \citenamefont
  {Conwell}(1954)}]{Debye1954}%
  \BibitemOpen
  \bibfield  {author} {\bibinfo {author} {\bibfnamefont {P.~P.}\ \bibnamefont
  {Debye}}\ and\ \bibinfo {author} {\bibfnamefont {E.~M.}\ \bibnamefont
  {Conwell}},\ }\bibfield  {title} {\bibinfo {title} {Electrical properties
  n-type germanium},\ }\href {https://doi.org/10.1103/physrev.93.693}
  {\bibfield  {journal} {\bibinfo  {journal} {Physical Review}\ }\textbf
  {\bibinfo {volume} {93}},\ \bibinfo {pages} {693–706} (\bibinfo {year}
  {1954})}\BibitemShut {NoStop}%
\bibitem [{\citenamefont {Spitzer}\ \emph {et~al.}(1961)\citenamefont
  {Spitzer}, \citenamefont {Trumbore},\ and\ \citenamefont
  {Logan}}]{Spitzer1961}%
  \BibitemOpen
  \bibfield  {author} {\bibinfo {author} {\bibfnamefont {W.~G.}\ \bibnamefont
  {Spitzer}}, \bibinfo {author} {\bibfnamefont {F.~A.}\ \bibnamefont
  {Trumbore}},\ and\ \bibinfo {author} {\bibfnamefont {R.~A.}\ \bibnamefont
  {Logan}},\ }\bibfield  {title} {\bibinfo {title} {Properties of heavily doped
  n-type germanium},\ }\href {https://doi.org/10.1063/1.1728243} {\bibfield
  {journal} {\bibinfo  {journal} {Journal of Applied Physics}\ }\textbf
  {\bibinfo {volume} {32}},\ \bibinfo {pages} {1822–1830} (\bibinfo {year}
  {1961})}\BibitemShut {NoStop}%
\bibitem [{\citenamefont {van Setten}\ \emph {et~al.}(2018)\citenamefont {van
  Setten}, \citenamefont {Giantomassi}, \citenamefont {Bousquet}, \citenamefont
  {Verstraete}, \citenamefont {Hamann}, \citenamefont {Gonze},\ and\
  \citenamefont {Rignanese}}]{vanSetten2018}%
  \BibitemOpen
  \bibfield  {author} {\bibinfo {author} {\bibfnamefont {M.}~\bibnamefont {van
  Setten}}, \bibinfo {author} {\bibfnamefont {M.}~\bibnamefont {Giantomassi}},
  \bibinfo {author} {\bibfnamefont {E.}~\bibnamefont {Bousquet}}, \bibinfo
  {author} {\bibfnamefont {M.}~\bibnamefont {Verstraete}}, \bibinfo {author}
  {\bibfnamefont {D.}~\bibnamefont {Hamann}}, \bibinfo {author} {\bibfnamefont
  {X.}~\bibnamefont {Gonze}},\ and\ \bibinfo {author} {\bibfnamefont {G.-M.}\
  \bibnamefont {Rignanese}},\ }\bibfield  {title} {\bibinfo {title} {The
  pseudodojo: Training and grading a 85 element optimized norm-conserving
  pseudopotential table},\ }\href {https://doi.org/10.1016/j.cpc.2018.01.012}
  {\bibfield  {journal} {\bibinfo  {journal} {Computer Physics Communications}\
  }\textbf {\bibinfo {volume} {226}},\ \bibinfo {pages} {39–54} (\bibinfo
  {year} {2018})}\BibitemShut {NoStop}%
\bibitem [{\citenamefont {Perdew}\ \emph {et~al.}(1996)\citenamefont {Perdew},
  \citenamefont {Burke},\ and\ \citenamefont {Ernzerhof}}]{Perdew1996}%
  \BibitemOpen
  \bibfield  {author} {\bibinfo {author} {\bibfnamefont {J.~P.}\ \bibnamefont
  {Perdew}}, \bibinfo {author} {\bibfnamefont {K.}~\bibnamefont {Burke}},\ and\
  \bibinfo {author} {\bibfnamefont {M.}~\bibnamefont {Ernzerhof}},\ }\bibfield
  {title} {\bibinfo {title} {Generalized gradient approximation made simple},\
  }\href {https://doi.org/10.1103/physrevlett.77.3865} {\bibfield  {journal}
  {\bibinfo  {journal} {Physical Review Letters}\ }\textbf {\bibinfo {volume}
  {77}},\ \bibinfo {pages} {3865–3868} (\bibinfo {year} {1996})}\BibitemShut
  {NoStop}%
\bibitem [{\citenamefont {Tran}\ and\ \citenamefont {Blaha}(2009)}]{Tran2009}%
  \BibitemOpen
  \bibfield  {author} {\bibinfo {author} {\bibfnamefont {F.}~\bibnamefont
  {Tran}}\ and\ \bibinfo {author} {\bibfnamefont {P.}~\bibnamefont {Blaha}},\
  }\bibfield  {title} {\bibinfo {title} {Accurate band gaps of semiconductors
  and insulators with a semilocal exchange-correlation potential},\ }\bibfield
  {journal} {\bibinfo  {journal} {Physical Review Letters}\ }\textbf {\bibinfo
  {volume} {102}},\ \href {https://doi.org/10.1103/physrevlett.102.226401}
  {10.1103/physrevlett.102.226401} (\bibinfo {year} {2009})\BibitemShut
  {NoStop}%
\bibitem [{\citenamefont {Perdew}\ and\ \citenamefont
  {Zunger}(1981)}]{Perdew1981}%
  \BibitemOpen
  \bibfield  {author} {\bibinfo {author} {\bibfnamefont {J.~P.}\ \bibnamefont
  {Perdew}}\ and\ \bibinfo {author} {\bibfnamefont {A.}~\bibnamefont
  {Zunger}},\ }\bibfield  {title} {\bibinfo {title} {Self-interaction
  correction to density-functional approximations for many-electron systems},\
  }\href {https://doi.org/10.1103/physrevb.23.5048} {\bibfield  {journal}
  {\bibinfo  {journal} {Physical Review B}\ }\textbf {\bibinfo {volume} {23}},\
  \bibinfo {pages} {5048–5079} (\bibinfo {year} {1981})}\BibitemShut
  {NoStop}%
\bibitem [{\citenamefont {Perdew}\ and\ \citenamefont
  {Wang}(1992)}]{Perdew1992}%
  \BibitemOpen
  \bibfield  {author} {\bibinfo {author} {\bibfnamefont {J.~P.}\ \bibnamefont
  {Perdew}}\ and\ \bibinfo {author} {\bibfnamefont {Y.}~\bibnamefont {Wang}},\
  }\bibfield  {title} {\bibinfo {title} {Accurate and simple analytic
  representation of the electron-gas correlation energy},\ }\href
  {https://doi.org/10.1103/physrevb.45.13244} {\bibfield  {journal} {\bibinfo
  {journal} {Physical Review B}\ }\textbf {\bibinfo {volume} {45}},\ \bibinfo
  {pages} {13244–13249} (\bibinfo {year} {1992})}\BibitemShut {NoStop}%
\bibitem [{\citenamefont {Collings}(1980)}]{Collings1980}%
  \BibitemOpen
  \bibfield  {author} {\bibinfo {author} {\bibfnamefont {P.~J.}\ \bibnamefont
  {Collings}},\ }\bibfield  {title} {\bibinfo {title} {Simple measurement of
  the band gap in silicon and germanium},\ }\href
  {https://doi.org/10.1119/1.12172} {\bibfield  {journal} {\bibinfo  {journal}
  {American Journal of Physics}\ }\textbf {\bibinfo {volume} {48}},\ \bibinfo
  {pages} {197–199} (\bibinfo {year} {1980})}\BibitemShut {NoStop}%
\bibitem [{\citenamefont {Sukheeja}(1983)}]{Sukheeja1983}%
  \BibitemOpen
  \bibfield  {author} {\bibinfo {author} {\bibfnamefont {B.~D.}\ \bibnamefont
  {Sukheeja}},\ }\bibfield  {title} {\bibinfo {title} {Measurement of the band
  gap in silicon and germanium},\ }\href {https://doi.org/10.1119/1.13436}
  {\bibfield  {journal} {\bibinfo  {journal} {American Journal of Physics}\
  }\textbf {\bibinfo {volume} {51}},\ \bibinfo {pages} {72–72} (\bibinfo
  {year} {1983})}\BibitemShut {NoStop}%
\bibitem [{\citenamefont {Green}(1990)}]{green1990intrinsic}%
  \BibitemOpen
  \bibfield  {author} {\bibinfo {author} {\bibfnamefont {M.~A.}\ \bibnamefont
  {Green}},\ }\bibfield  {title} {\bibinfo {title} {Intrinsic concentration,
  effective densities of states, and effective mass in silicon},\ }\href
  {https://doi.org/10.1063/1.345414} {\bibfield  {journal} {\bibinfo  {journal}
  {Journal of Applied Physics}\ }\textbf {\bibinfo {volume} {67}},\ \bibinfo
  {pages} {2944} (\bibinfo {year} {1990})}\BibitemShut {NoStop}%
\bibitem [{\citenamefont {Bludau}\ \emph {et~al.}(1974)\citenamefont {Bludau},
  \citenamefont {Onton},\ and\ \citenamefont {Heinke}}]{bludau1974temperature}%
  \BibitemOpen
  \bibfield  {author} {\bibinfo {author} {\bibfnamefont {W.}~\bibnamefont
  {Bludau}}, \bibinfo {author} {\bibfnamefont {A.}~\bibnamefont {Onton}},\ and\
  \bibinfo {author} {\bibfnamefont {W.}~\bibnamefont {Heinke}},\ }\bibfield
  {title} {\bibinfo {title} {Temperature dependence of the band gap of silicon
  and germanium},\ }\href {https://doi.org/10.1063/1.1663436} {\bibfield
  {journal} {\bibinfo  {journal} {Journal of Applied Physics}\ }\textbf
  {\bibinfo {volume} {45}},\ \bibinfo {pages} {1846} (\bibinfo {year}
  {1974})}\BibitemShut {NoStop}%
\bibitem [{\citenamefont {Madelung}(2004)}]{Madelung2004}%
  \BibitemOpen
  \bibfield  {author} {\bibinfo {author} {\bibfnamefont {O.}~\bibnamefont
  {Madelung}},\ }\href {https://doi.org/10.1007/978-3-642-18865-7} {\emph
  {\bibinfo {title} {Semiconductors: Data Handbook}}}\ (\bibinfo  {publisher}
  {Springer Berlin Heidelberg},\ \bibinfo {year} {2004})\BibitemShut {NoStop}%
\bibitem [{\citenamefont {Gonze}\ \emph {et~al.}(2010)\citenamefont {Gonze},
  \citenamefont {Boulanger},\ and\ \citenamefont {C\^oté}}]{Gonze2010}%
  \BibitemOpen
  \bibfield  {author} {\bibinfo {author} {\bibfnamefont {X.}~\bibnamefont
  {Gonze}}, \bibinfo {author} {\bibfnamefont {P.}~\bibnamefont {Boulanger}},\
  and\ \bibinfo {author} {\bibfnamefont {M.}~\bibnamefont {C\^oté}},\
  }\bibfield  {title} {\bibinfo {title} {Theoretical approaches to the
  temperature and zero‐point motion effects on the electronic band
  structure},\ }\href {https://doi.org/10.1002/andp.201000100} {\bibfield
  {journal} {\bibinfo  {journal} {Annalen der Physik}\ }\textbf {\bibinfo
  {volume} {523}},\ \bibinfo {pages} {168–178} (\bibinfo {year}
  {2010})}\BibitemShut {NoStop}%
\bibitem [{\citenamefont {Poncé}\ \emph {et~al.}(2015)\citenamefont {Poncé},
  \citenamefont {Gillet}, \citenamefont {Laflamme~Janssen}, \citenamefont
  {Marini}, \citenamefont {Verstraete},\ and\ \citenamefont
  {Gonze}}]{ponce2015}%
  \BibitemOpen
  \bibfield  {author} {\bibinfo {author} {\bibfnamefont {S.}~\bibnamefont
  {Poncé}}, \bibinfo {author} {\bibfnamefont {Y.}~\bibnamefont {Gillet}},
  \bibinfo {author} {\bibfnamefont {J.}~\bibnamefont {Laflamme~Janssen}},
  \bibinfo {author} {\bibfnamefont {A.}~\bibnamefont {Marini}}, \bibinfo
  {author} {\bibfnamefont {M.}~\bibnamefont {Verstraete}},\ and\ \bibinfo
  {author} {\bibfnamefont {X.}~\bibnamefont {Gonze}},\ }\bibfield  {title}
  {\bibinfo {title} {Temperature dependence of the electronic structure of
  semiconductors and insulators},\ }\href {https://doi.org/10.1063/1.4927081}
  {\bibfield  {journal} {\bibinfo  {journal} {The Journal of Chemical Physics}\
  }\textbf {\bibinfo {volume} {143}},\ \bibinfo {pages} {102813} (\bibinfo
  {year} {2015})},\ \Eprint
  {https://arxiv.org/abs/https://pubs.aip.org/aip/jcp/article-pdf/doi/10.1063/1.4927081/15502798/102813\_1\_online.pdf}
  {https://pubs.aip.org/aip/jcp/article-pdf/doi/10.1063/1.4927081/15502798/102813\_1\_online.pdf}
  \BibitemShut {NoStop}%
\end{thebibliography}%

\clearpage

\appendix
\renewcommand{\thefigure}{B\arabic{figure}}
\setcounter{figure}{0}
\section{Computational details}\label{appendix:computational_detail}

All electronic structure calculations were performed using \textsc{ABINIT}\cite{Verstraete2025,Gonze2020,Romero2020,vanSetten2018} with a plane-wave basis. For silicon, we employed the Generalized Gradient Approximation (GGA) in the Perdew-Burke-Ernzerhof (PBE) formulation~\cite{Perdew1996}, while for germanium we used the Becke-Johnson (BJ) exchange potential~\cite{Tran2009} combined with local-density approximation (LDA) correlation.\cite{Perdew1981,Perdew1992} The latter choice was necessary because standard LDA or GGA severely underestimates the Ge band gap, often predicting a metallic ground state, whereas BJ+LDA yields a finite indirect gap in close agreement with experiment. With these functionals, the calculated indirect band gaps are $0.52~\text{eV}$ for Si and $0.686~\text{eV}$ for Ge, compared to the experimental values\cite{Collings1980,Sukheeja1983} of $1.17~\text{eV}$~\cite{green1990intrinsic} and $0.74~\text{eV}$~\cite{bludau1974temperature} at $0~\text{K}$. The plane-wave kinetic-energy cutoff was set to $40~\text{Ha}$. Both materials adopt the diamond cubic structure with two atoms per primitive cell, with optimized lattice constants $a_{\mathrm{Si}} = 5.440~\text{\AA}$ and $a_{\mathrm{Ge}} = 5.674~\text{\AA}$, in close agreement with the experimental values of $5.431~\text{\AA}$ and $5.658~\text{\AA}$.~\cite{Madelung2004} Brillouin-zone integrations for both electronic states and phonons were performed on uniform $8\times 8\times 8$ $k$- and $q$-point grids within DFPT. The resulting e-ph matrix elements were then used in MBPT to evaluate the e-ph self-energies and further spectral functions.

\section{ Convergence studies}\label{appendix:convergence study}

\begin{figure}
	\centering
	\includegraphics[width = 9.5 cm]{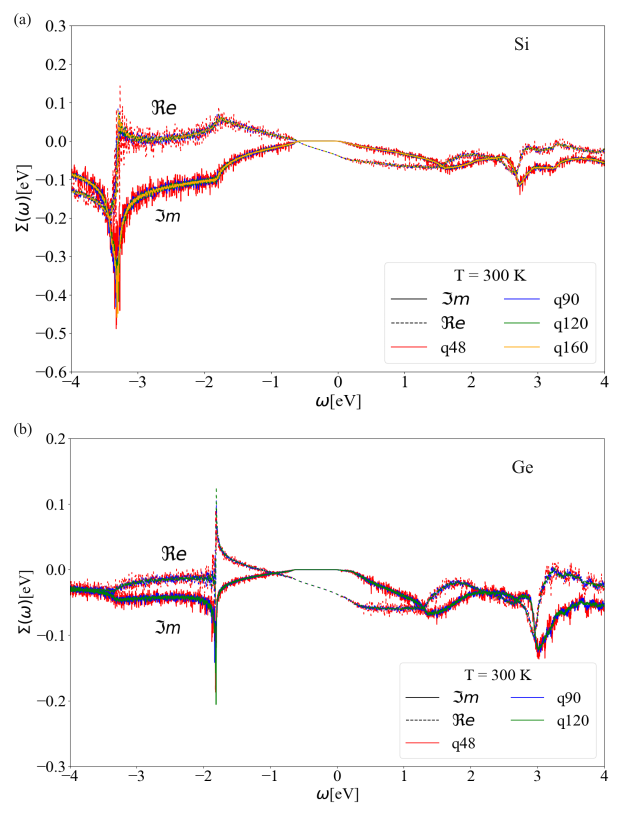}
	\caption{Real (\(\Re e\)) and imaginary (\(\Im m\)) parts of the electron self-energy at the CBM for (a) Si and (b) Ge at \(T = 300~\mathrm{K}\), calculated within the DM approach using \(\eta = 1~\mathrm{meV}\). Results are shown for different \(N\) grids: \(N = 48, 90, 120\) for both Si and Ge, and additionally \(N = 160\) for Si. The self-energy is smooth and numerically converged at \(N = 160\) for Si and \(N = 120\) for Ge.}
	\label{fig:appendix_self_energy_si_ge}
\end{figure}

We also examine the convergence behavior of the real and imaginary parts of the electron–phonon self-energy as a function of the \( q \)-point sampling density \( N \) and the broadening parameter \( \eta \). This analysis is essential, as the CE method is highly sensitive to the quality of the input Fan–Migdal self-energy. Our numerical tests reveal that the real part of the self-energy at the Kohn–Sham (KS) energy converges adequately with relatively coarse \( q \)-point grids and larger values of \( \eta \), provided that a sufficient number of empty electronic states are included. However, the imaginary part of the self-energy demands significantly denser \( q \)-point sampling and smaller values of \( \eta \) for convergence. We perform a single, fully consistent calculation for both components, ensuring that the \( q \)-mesh is chosen based on the stricter convergence requirements of the imaginary part. The resulting improvement in both the real and imaginary parts of the self-energy with increasing \( q \)-point density is shown in \cref{fig:appendix_self_energy_si_ge}, where the noise level decreases systematically as the sampling becomes denser. For subsequent calculations of spectral functions within the DM and CE approaches, we adopt a \( q \)-mesh of \( N = 160 \) for silicon and \( N = 120 \) for germanium, which achieves low computational noise and convergence errors of only \( 0.4\% \) for Si and \( 0.5\% \) for Ge for the self-energy at the CBM. However, for mobility calculations, the transport energy window must be explicitly resolved, and a direct evaluation on such dense grids is computationally prohibitive; in this case, we employ the double-grid method detailed in \cref{appendix:double_grid}.

An additional observation is the linear deviation of \( \mathrm{Im}\,\Sigma_{\mathrm{CBM}}(\omega) \) at \( \omega_{\mathrm{LO}} \) from its value at \( \epsilon_{\mathrm{CBM}} \). This behavior contrasts with that in the Fr\"ohlich model and polar materials, where the e–ph matrix elements diverge for small \( q \), leading to a pronounced peak in the self-energy at the LO phonon frequency.

\begin{figure}
	\centering
	\includegraphics[width = 9.5 cm]{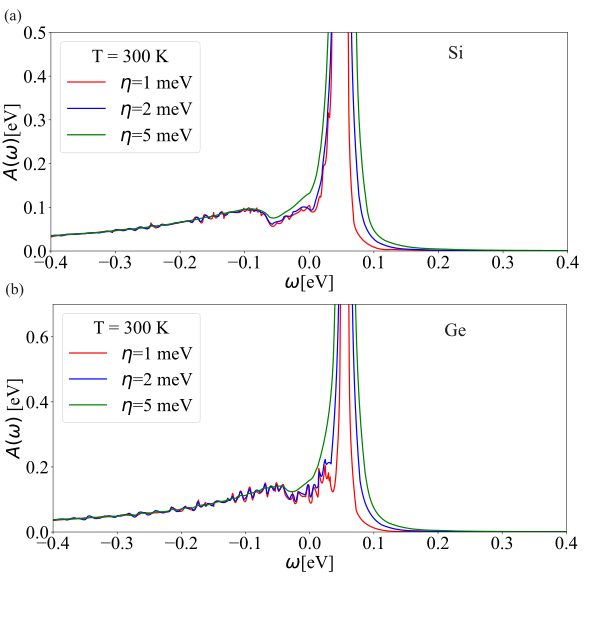}
	\caption{Spectral function at the VBM for (a) Si and (b) Ge at \(T = 300~\mathrm{K}\), computed within the DM formalism using a \(N = 160\) for Si and \(N = 120\) for Ge, for different values of \(\eta = 1, 2, 5~\mathrm{meV}\).
}
	\label{fig:appendix_dm_spectra_si_ge_eta}
\end{figure}

For a fixed \( q \)-mesh size, we also examined the effect of $\eta$ on the DM spectral function at both the CBM and VBM. Figure~\ref{fig:appendix_dm_spectra_si_ge_eta} shows the results for the VBM in Si and Ge. At \(\eta = 1~\mathrm{meV}\), the QP peak remains sharp and satellite features are well resolved, with only minor oscillations in the spectrum of both the materials attributable to residual numerical noise. Increasing \(\eta\) to \(2~\mathrm{meV}\) reduces these oscillations but introduces noticeable artificial broadening of the QP peak, while \(\eta = 5~\mathrm{meV}\) further suppresses fine spectral details. Based on these observations, we adopt \(\eta = 1~\mathrm{meV}\) for both Si and Ge, as it offers the best compromise between preserving intrinsic spectral resolution and maintaining numerical stability.

Another important aspect of convergence in the self-energy calculations is the summation over the band index \( m \) in \cref{eq:3_Fan_self_energy}, which includes contributions from unoccupied electronic states. Achieving numerical convergence for these empty states can be computationally challenging. A practical alternative is to replace high-energy bands with solutions obtained from a non-self-consistent Sternheimer equation as done in Ref \cite{Gonze2010, abreu2022}.
\renewcommand{\thefigure}{C\arabic{figure}}
\setcounter{figure}{0}
\section{Cumulant functions}\label{appendix:cumulant_function}

\begin{figure}[h]
	\centering
	\includegraphics[width = 9.5 cm]{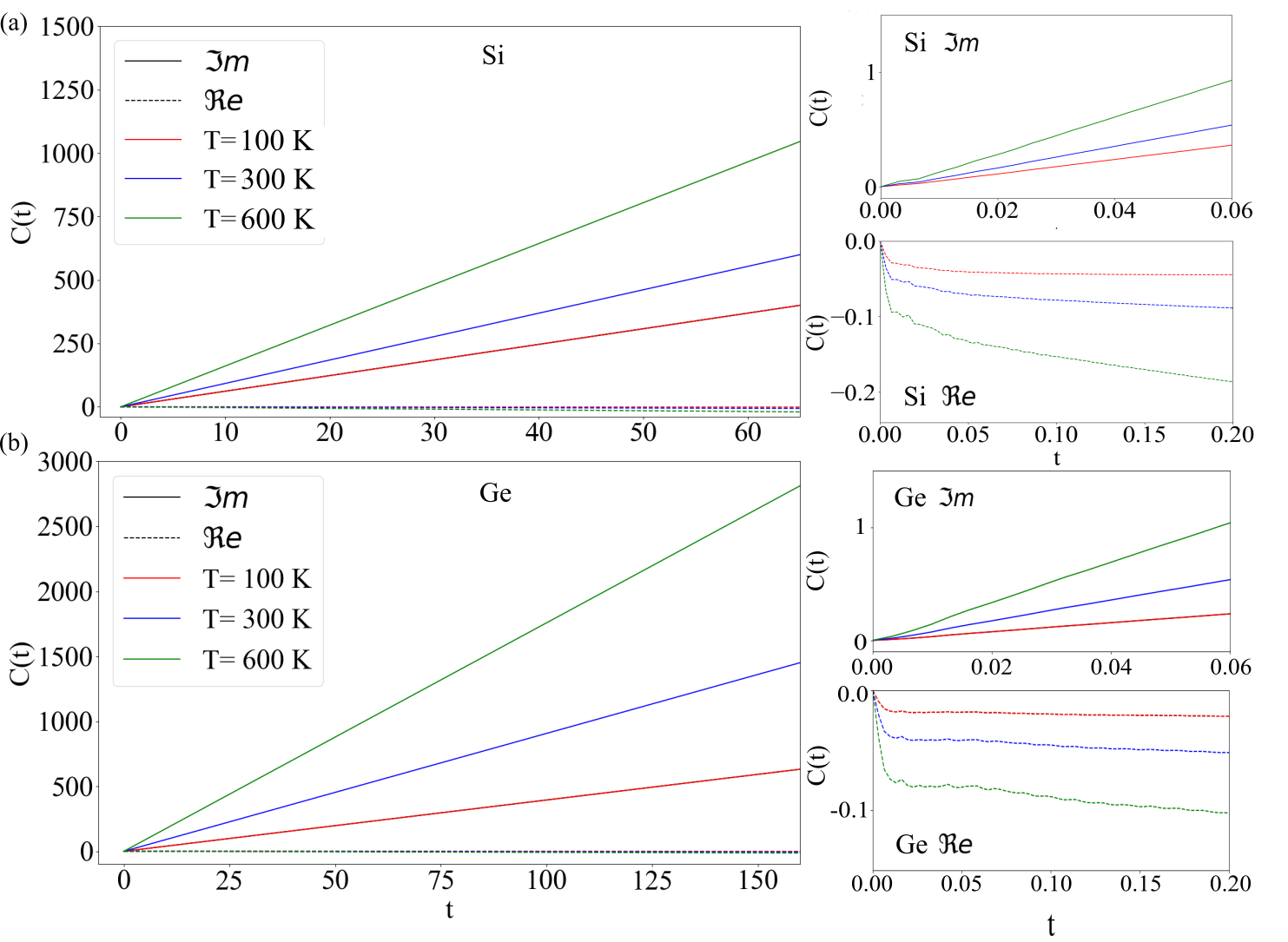}
	\caption{ Cumulant function \( C(t) \) at the CBM for (a) Si (0–65 ps) and (b) Ge (0–160 ps) at temperatures of 100, 300, and 600~K. Solid lines denote the imaginary part \(\Im m\ C(t)\), and dashed lines denote the real part \(\Re e\ C(t)\). The magnitude of \(\Im m\ C(t)\) increases linearly with temperature, while \(\Re e\ C(t)\) remains small in magnitude but exhibits a clear non-linear variation at short times (\(t \to 0\)), as highlighted in the right-hand panels.
}
	\label{fig:appendix_cumulant_si_ge}
\end{figure}

\cref{fig:appendix_cumulant_si_ge} illustrates the cumulant function at temperatures \( T = 100 \), \( 300 \), and \( 600 \, \text{K} \) for both silicon and germanium. Both the real and imaginary parts exhibit linear behavior at large times. The slope of the real part at long times corresponds to \( \Im m\Sigma_{\text{CBM}}(\epsilon_{\text{CBM}}^0) \), while the slope of the imaginary part corresponds to \( \Re e\Sigma_{\text{CBM}}(\epsilon_{\text{CBM}}^0)\). The intercept of \(\Re eC(t), w_k\) arises from the asymmetry in the denominator of \( \Sigma \) when summing over all \( q \)-points, as described in \cref{eq:11_cumulant_t_inf}. The growth of \(\Re e\ C(t) \) as \( t \to \infty \) contributes to the accelerated decay of the QP, ultimately determining its lifetime, given by \( \tau = 1 / (2\Gamma) \).

 The increase in $|\Re e\,C(t)|$ with temperature corresponds to larger $\Gamma$ (shorter lifetimes), whereas the steeper $\Im m\,C(t)$ indicates a larger quasiparticle energy shift $\Lambda$. Comparing the two materials, the magnitude of the long-time slope of $\Re e\,C(t)$ is smaller in Ge than in Si at the same temperature, implying $\Gamma_{\mathrm{Ge}}<\Gamma_{\mathrm{Si}}$ and therefore longer CBM lifetimes in Ge—consistent with its higher electron mobility. The panels on the right zoom into the short-time behavior of $\Re e\,C(t)$ as $t\!\to\!0$, where deviations from linearity encode multi-phonon processes responsible for generating satellite features in the spectral function.

\renewcommand{\thefigure}{D\arabic{figure}}
\setcounter{figure}{0}

\section{Double-grid approximation for mobility calculations}\label{appendix:double_grid}
To compute the mobilities using the Kubo-Greenwood formalism requiring dense $k$- and $q$-meshes (e.g., $N = 160$ or $N = 120$ for Si and Ge respectively), a direct calculation becomes computationally prohibitive due to the large memory requirements and the cost of evaluating the e-ph matrix elements $g_{mn\nu}(\mathbf{k},\mathbf{q})$. To address this, we employed the double-grid (DG) technique~\cite{Brunin2020prb}, in which a coarse mesh is used to evaluate the e-ph matrix elements, while a finer mesh is used to interpolate the energy-conserving Dirac $\delta$-functions associated with phonon absorption and emission. This approach is motivated by the observation that, apart from the Fr\"ohlich divergence in polar semiconductors\cite{ponce2015}, $g_{mn\nu}$ varies smoothly compared to the sharp structure of the $\delta$-functions.\cite{Brunin2020prb} The computationally less demanding quantities, $\epsilon_{n\mathbf{k}}$ and $\omega_{\mathbf{q}\nu}$, can thus be computed on a denser grid, significantly improving convergence without incurring the full cost of dense $g_{mn\nu}$ evaluations. For polar materials (not Si or Ge) in the vicinity of $\Gamma$ the Fr\"ohlich divergence occurs, and our algorithm treats the singularity by performing an angular numerical integration of the analytical form of the divergence in a small spherical region, controlled by the \texttt{eph\_frohl\_ntheta} parameter in ABINIT for the density of the angular integration. For dense $q$ grids the mini zone around $\Gamma$ is quite small and the neighboring $g_{mn\nu}$ still vary quite steeply. We have found that a simple linear interpolation of the matrix elements improves the convergence, though it can not handle band (anti)crossings or other complications.

\begin{figure}
	\centering
	\includegraphics[width = 9.5 cm]{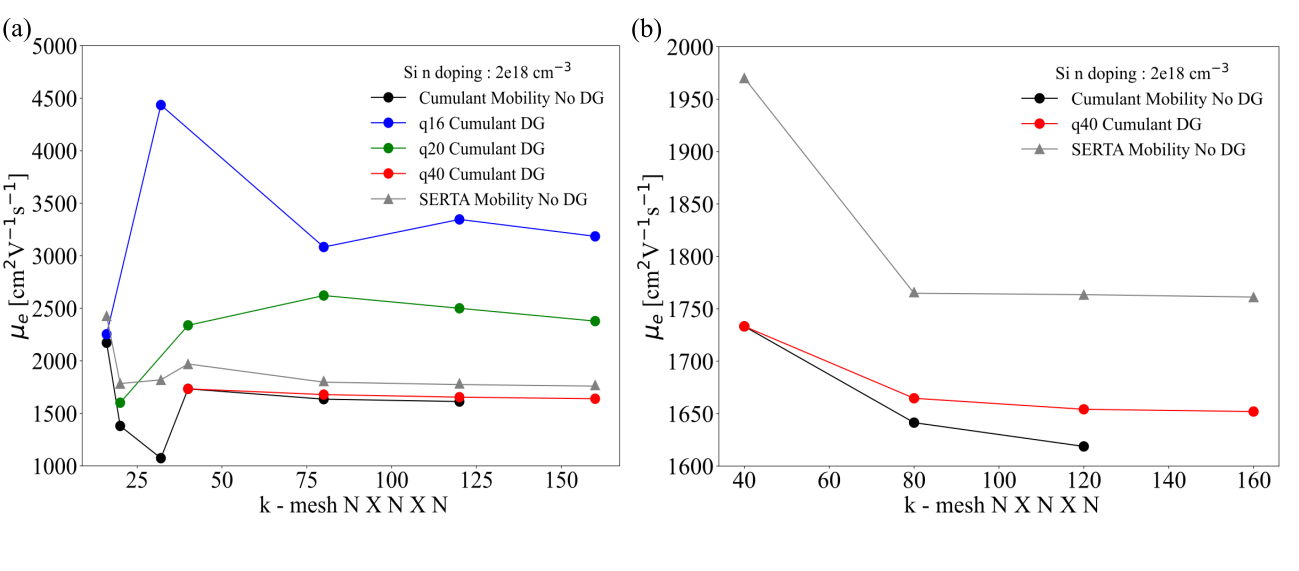}
\caption{%
Convergence of the electron mobility $\mu_{e}$ for $n$-type Si at a doping concentration of 
$2\times 10^{18}~\mathrm{cm}^{-3}$ using the double-grid (DG) technique. The fine $k$-mesh is denoted 
as $N\times N\times N$ (horizontal axis). 
(a) Mobilities computed using the cumulant approach without DG (black circles), the SERTA approach without DG (gray triangles), and cumulant calculations incorporating the DG technique with coarse $q$-meshes of 16 (blue), 20 (green), and 40 (red). 
(b) Zoomed view of the converged region near $1600$--$2000~\mathrm{cm^{2}V^{-1}s^{-1}}$. 
The $q40$ DG results converge faster and at lower computational cost than the no-DG cumulant, which remains unconverged even up to a $N=120$ mesh.
}
\label{fig:appendix_DG_Si}
\end{figure}

\cref{fig:appendix_DG_Si} shows the convergence of the electron mobility \(\mu_{e}\) for \(n\)-type silicon at a doping concentration of \(2\times10^{18}\,\mathrm{cm^{-3}}\), computed using the DG approach. The horizontal axis denotes the fine \(k\)-mesh \(N\times N\times N\). Black circles and gray triangles correspond to mobility values obtained without the DG approximation, using the cumulant and SERTA approaches, respectively. Colored symbols represent DG calculations employing coarse $q$-meshes of 16 (blue), 20 (green) and 40 (red). The blue and green DG curves exhibit large deviations from the converged values, which highlights the need for sufficient \(q\)-mesh resolution. By contrast, the \(q=40\) curve closely reproduces the dense-grid cumulant reference while incurring a much lower computational cost. \cref{fig:appendix_DG_Si} (b) gives a zoomed view of the converged window, where the \(q=40\) DG result attains convergence more rapidly and smoothly than the no-DG cumulant, which continues to drift even at a \(N=120\) mesh.  The SERTA mobility converges faster since it only involves the imaginary part of the electron self-energy and is computationally inexpensive. These tests demonstrate that the DG formalism accelerates numerical convergence while retaining the accuracy of full cumulant calculations, making it well suited for large-scale or high-throughput mobility calculations. The same study is performed for germanium. From this study we adopt a coarse grid of \(N_{\mathrm{coarse}}=40\) and fine grid of \(N_{\mathrm{fine}}=120\) in all subsequent mobility calculations for both Si and Ge.

\end{document}